\pdfoutput=1
\documentclass[letterpaper,twocolumn,10pt]{article}
\usepackage{usenix,epsfig,endnotes}
\usepackage{xspace}             
\usepackage{mdframed}    

\newcommand{\sysname}[0]{StingRay\xspace}

\newcommand{\papertitle}[0]{Technical Report: When Does Machine Learning FAIL? Generalized Transferability for Evasion and Poisoning Attacks\vspace{-0.3in}}

\usepackage{pifont} 
\newcommand{\cmark}{\ding{51}}
\newcommand{\xmark}{\ding{55}}

\usepackage{combelow}
\usepackage{amsmath}
\usepackage{amssymb}
\usepackage[normalem]{ulem}   
\usepackage{soul}       
\usepackage{xcolor}       
\usepackage{url}                  

\usepackage{subcaption}
\usepackage{calc}                 
\usepackage{shadow}             
\usepackage{fancyvrb}           
\usepackage{xspace}             
\usepackage{ifthen}               
\usepackage{comment}            
\usepackage{multirow}           
\usepackage{dcolumn}            
\usepackage{colortbl}
\usepackage{tabularx}           
\usepackage{mdwtab}             
\usepackage{mdwlist}      
\usepackage{ifpdf}                
\usepackage{rotating}           
\usepackage{tablefootnote}

\usepackage{listings}
\usepackage{wrapfig}            
\usepackage{pdfpages}     
\usepackage{graphicx}           
\usepackage{enumitem} 
\usepackage{longtable}

\usepackage{algorithm}
\usepackage[noend]{algpseudocode}
\usepackage{diagbox}
\usepackage{fdsymbol} 

\usepackage[table, subgroups]{svn-multi}
\usepackage{hyperref}

\svnidlong
{$HeadURL: https://mc2.umiacs.umd.edu/svn/SDSatUMD/papers/2016_ml_reputation/ml_reputation.tex $}
{$LastChangedDate: 2017-05-19 23:43:29 -0400 (Fri, 19 May 2017) $}
{$LastChangedRevision: 5031 $}
{$LastChangedBy: osuciu $}

\def\paperversionmajor{1}          
\def\paperversionminor{\svnrev}    

\def\monthName#1{\ifcase#1\or
  January\or February\or March\or April\or May\or June\or
  July\or August\or September\or October\or November\or December\fi}

\hypersetup{pdftitle={\papertitle}}
\hypersetup{pdfproducer={v. \paperversionmajor.\paperversionminor}}
\ifpdf
\fi

\usepackage[font=footnotesize]{caption}
\sethlcolor{lime}

\newcommand{\usenix}[1]{\ignorespaces}
\newcommand{\arxiv}[1]{#1}

\newcounter{hypothesis}                                     

\newcommand{\topic}[1]{\vspace{0.1in}\noindent\textbf{#1.}}

\excludecomment{RoniDetails}

\graphicspath{ {figures/} }

\renewcommand{\labelenumi}{\textbf{\Roman{enumi}}}
\newcommand{\RNum}[1]{\textbf{\uppercase\expandafter{\romannumeral #1\relax}}}

\begin{document}
\pagenumbering{gobble}
%
\title{\papertitle}


\author{
{\rm Octavian Suciu}\\
\and
{\rm Radu M\u{a}rginean}\\
\and
{\rm Yi\u{g}itcan Kaya}\\
\and
{\rm Hal Daum\'e III}\\
\and
{\rm Tudor Dumitra\cb{s}}\\
University of Maryland, College Park
} 

%

\maketitle

%
%

\begin{abstract}

Recent results suggest that 
attacks against supervised machine learning systems 
are
quite effective, while defenses are easily bypassed by new attacks. 
%
%
However, the specifications for machine learning systems currently lack precise adversary definitions, and the existing 
attacks make diverse, potentially unrealistic assumptions about the strength of the adversary who launches them.
%
%
We propose the \textbf{FAIL} attacker model, which describes the adversary's knowledge and control along four dimensions.
%
Our model allows us to consider a wide range of weaker adversaries 
who have limited control and incomplete knowledge of the features, learning algorithms and training instances utilized. 
%

To evaluate the utility of the \textbf{FAIL} model, we consider the problem of conducting targeted poisoning attacks in a realistic setting:
the crafted poison samples must have clean labels, must be individually and collectively inconspicuous, and
%
must exhibit a generalized form of transferability, 
defined by the \textbf{FAIL} model.
By taking these constraints into account, we design \sysname, a targeted poisoning attack that is practical against 4 machine learning applications, which use 3 different learning algorithms,
and can bypass 2 existing defenses.
Conversely, we show that a prior evasion attack is less effective under generalized transferability.
%
Such attack evaluations, under the \textbf{FAIL} adversary model, may also suggest
promising directions for 
future defenses.

\end{abstract}

%
%

\section{Introduction}
\label{sec:intro}

\begin{figure*}[t]

    \centering
    \begin{subfigure}{.24\textwidth}
        \centering
        \includegraphics[height=0.93in]{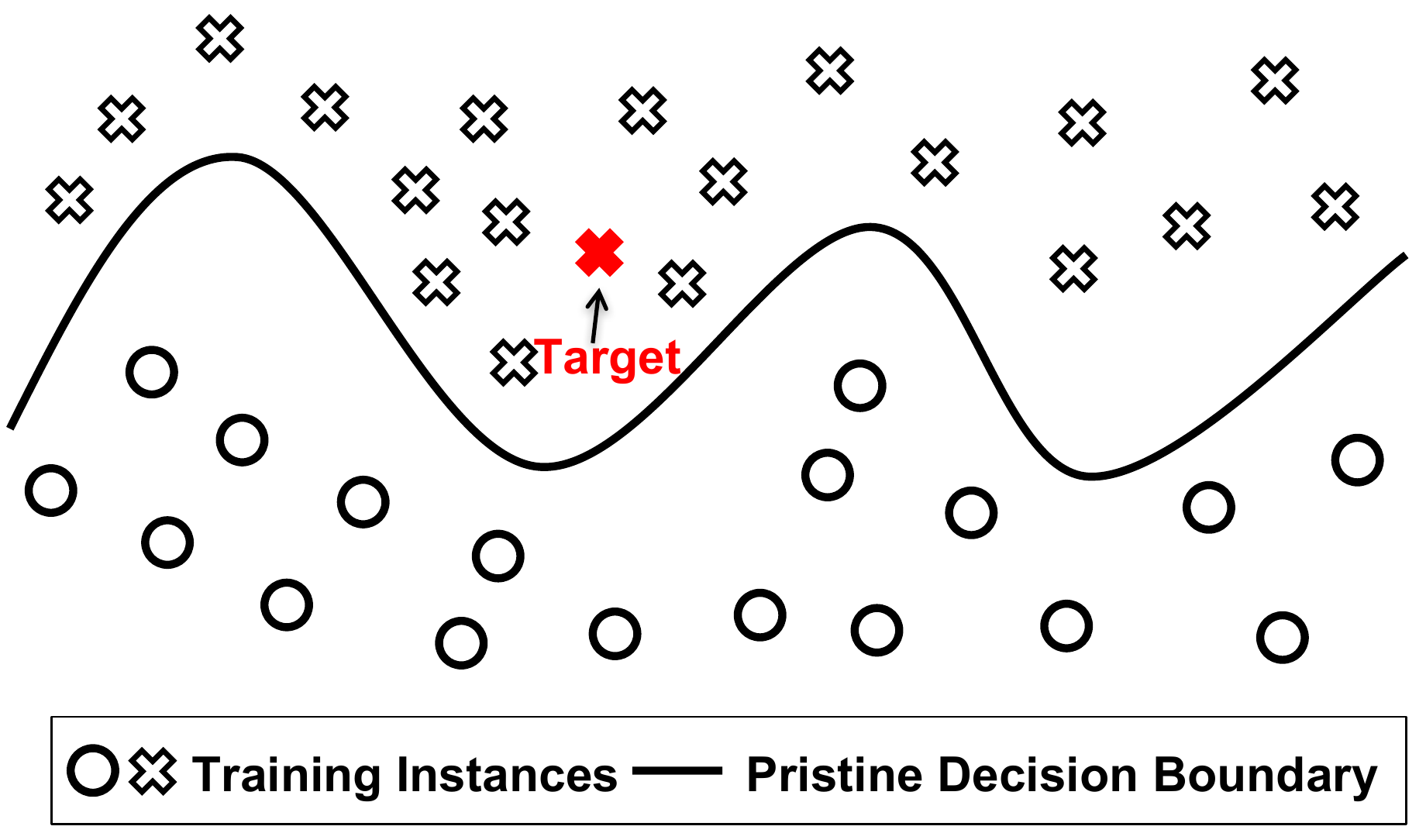}
        \vspace{-5pt}
        \caption{}
        \label{fig:attack_illustration_pristine}
    \end{subfigure}%
    \begin{subfigure}{.24\textwidth}
        \centering
        \includegraphics[height=0.93in]{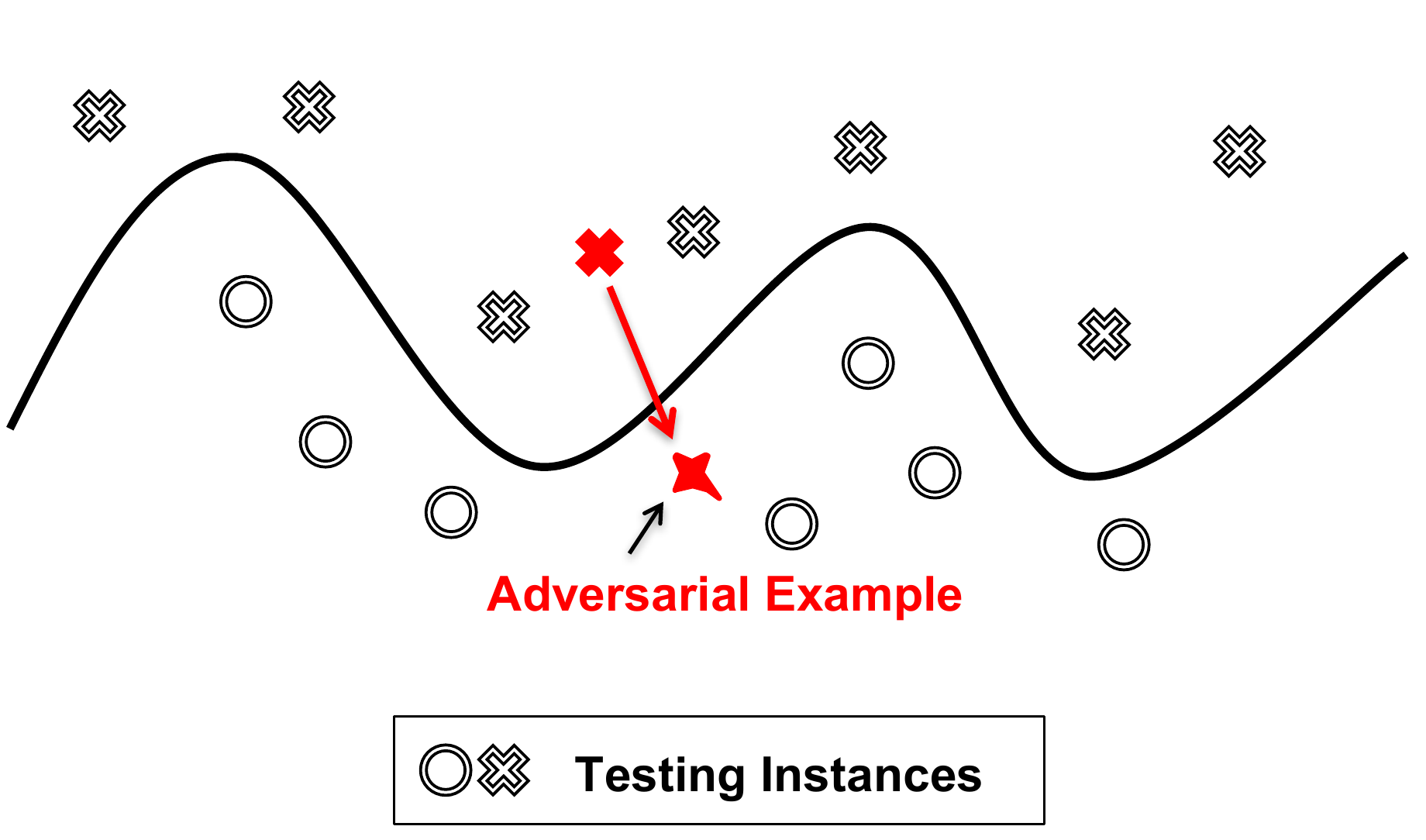}
        \vspace{-5pt}
        \caption{}
        \label{fig:attack_illustration_evasion}
    \end{subfigure}%
    \begin{subfigure}{.24\textwidth}
        \centering
        \includegraphics[height=0.93in]{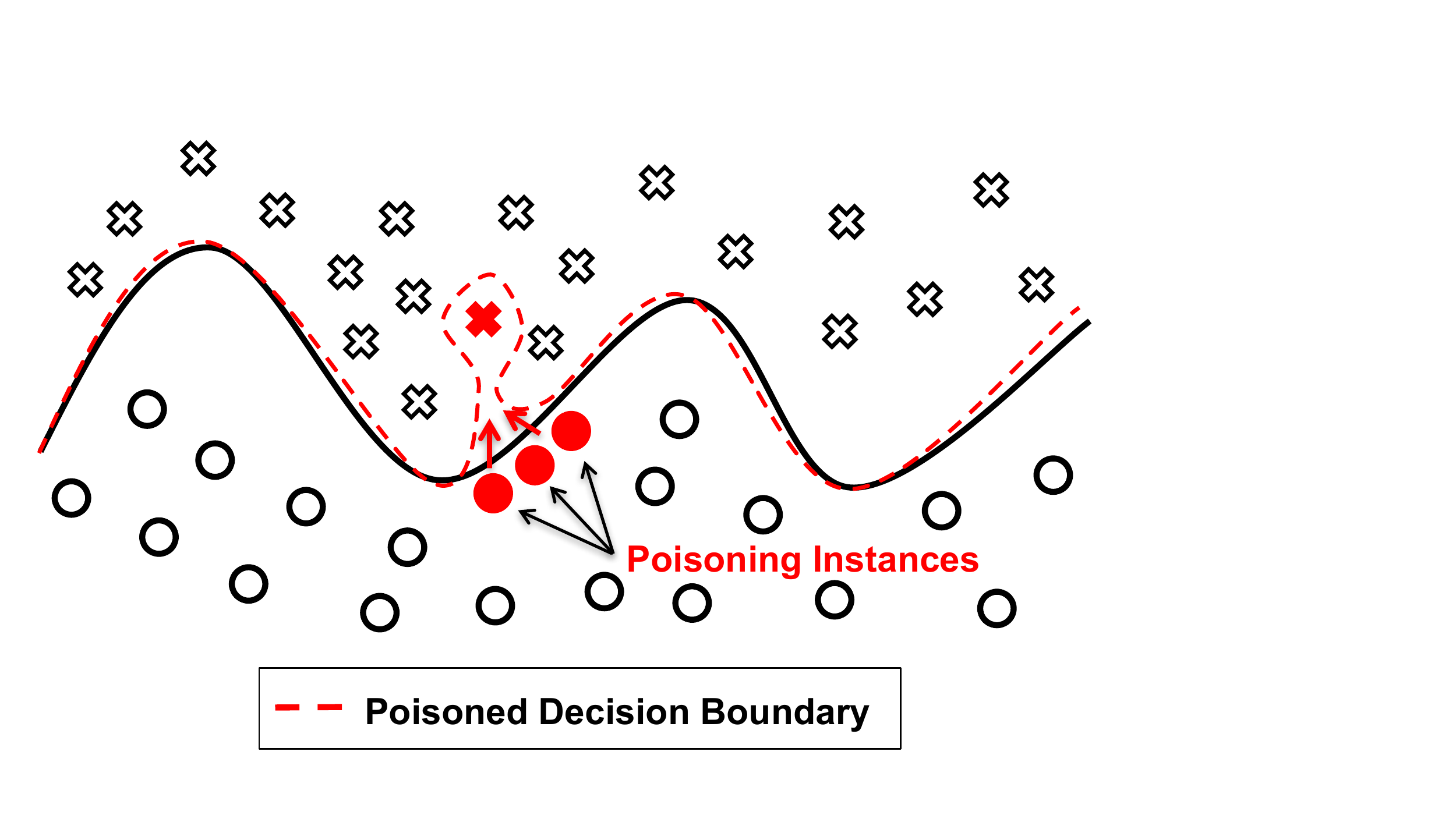}
        \vspace{-5pt}
        \caption{}
        \label{fig:attack_illustration_poisoning}
    \end{subfigure}
    \begin{subfigure}{.24\textwidth}
        \centering
        \includegraphics[height=0.93in]{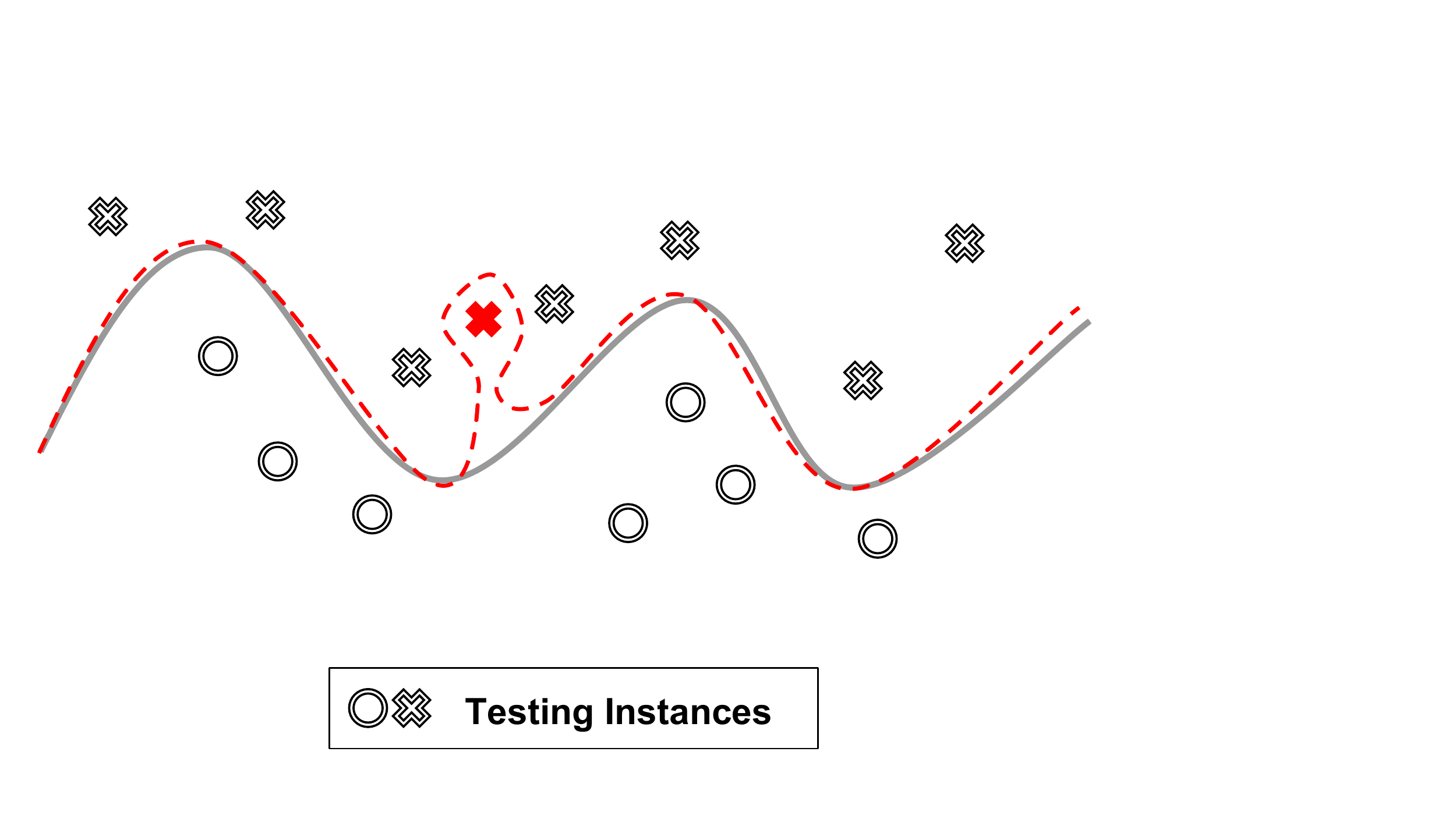}
        \vspace{-5pt}
        \caption{}
        \label{fig:attack_illustration_testing}
    \end{subfigure}
    \caption{Targeted attacks against machine learning classifiers. (a) The pristine classifier would correctly classify the target. (b) An evasion attack would modify the target to cross the decision boundary. (c) Correctly labeled poisoning instances change the learned decision boundary. (d) At testing time, the target is misclassified but other instances are correctly classified.}
    \label{fig:attack_illustration}
\end{figure*}

Machine learning (ML) systems are widely deployed in safety-critical domains that carry incentives for potential adversaries, such as finance~\cite{enstyoung}, medicine~\cite{googleresearchmedical}, the justice system~\cite{mittrjustice}, cybersecurity~\cite{kasperskylab}, or self-driving cars~\cite{bojarski2017explaining}.
An ML classifier automatically learns classification models using labeled observations (samples) from a \emph{training set}, without requiring predetermined rules for mapping inputs to labels.
It can then apply these models to predict labels for new samples in a \emph{testing set}.
%
An \emph{adversary} knows some or all of the ML system's parameters and uses this knowledge to craft training or testing samples that manipulate the decisions of the ML system according to the adversary's goal%
---for example, to avoid being sentenced by an ML-enhanced judge.

Recent work has focused primarily on
\emph{evasion} 
attacks~\cite{biggio2013evasion, szegedy2013intriguing,goodfellow2014explaining, xu2016automatically,papernot2016limitations,carlini2017towards},
which can induce a \emph{targeted} misclassification on a specific sample.
As illustrated in Figures~\ref{fig:attack_illustration_pristine} and~\ref{fig:attack_illustration_evasion}, these test time attacks work by mutating the target sample to push it across the model's decision boundary, without altering the training process or the decision boundary itself.
They are not applicable in situations where the adversary does not control the target sample---for example, when she aims to influence a malware detector to block a benign app developed by a competitor.
Prior research has also shown the feasibility of targeted \emph{poisoning} attacks~\cite{Nelson:2008:EML:1387709.1387716, mozaffari2015systematic}. 
As illustrated in Figure~\ref{fig:attack_illustration_poisoning}, these attacks usually blend crafted instances into the training set to push the model's boundary toward the target.
In consequence, they enable misclassifications for instances that the adversary cannot modify.

These attacks appear to be very effective, and the defenses proposed against them are often bypassed in follow-on work~\cite{carlini2017adversarial}.
However, to understand the actual security threat introduced by them, we must model the capabilities and limitations of realistic adversaries.
Evaluating poisoning and evasion attacks under assumptions that overestimate the capabilities of the adversary would lead to an inaccurate picture of the security threat posed to real-world applications.
For example, test time attacks often assume white-box access to the victim classifier ~\cite{carlini2017towards}. As most security-critical ML systems use proprietary models~\cite{kasperskylab}, these attacks might not reflect actual capabilities of a potential adversary.
Black-box attacks consider weaker adversaries, but they often make rigid assumptions about the adversary's knowledge when investigating the \emph{transferability} of an attack.
Transferability is a property of attack samples crafted locally, on a surrogate model that reflects the adversary's limited knowledge, allowing them to remain successful against the target model.
Specifically, black-box attacks often investigate transferability in the case where the local and target models use different training \emph{algorithms}~\cite{DBLP:journals/corr/PapernotMG16}.
In contrast, ML systems used in the security industry often resort to \emph{feature} secrecy (rather than algorithmic secrecy) to protect themselves against attacks, e.g. by incorporating undisclosed features 
for malware detection~\cite{Chau11:Polonium}.

In this paper, we 
make a first step towards modeling realistic adversaries who aim to conduct attacks against ML systems.
To this end, we propose the \textbf{FAIL} model, a general framework for the analysis of ML attacks in settings with a variable amount of adversarial knowledge and control over the victim, along four tunable dimensions: \textbf{F}eatures, \textbf{A}lgorithms, \textbf{I}nstances, and \textbf{L}everage.
By preventing any implicit assumptions about the adversarial capabilities, the model is able to accurately highlight the success rate of a wide range of attacks in realistic scenarios and forms a common ground for modeling adversaries.
Furthermore, the \textbf{FAIL} framework generalizes the transferability of attacks by providing a multidimensional basis for surrogate models.
This provides insights into the constraints of realistic adversaries, which could be explored in future research on defenses against these attacks.
For example, our evaluation suggests that crafting transferable samples with an existing evasion attack is more challenging than previously believed.

To evaluate the utility of the \textbf{FAIL} model, we consider the problem of \emph{conducting targeted poisoning attacks in a realistic setting}.
Specifically, we impose four constraints on the adversary. 
First, the poison samples must have \emph{clean labels}, as the adversary can inject them into the training set of the model under attack but cannot determine how they are labeled. 
Second, the samples must be \emph{individually inconspicuous}, i.e. to be very similar to the existing training instances in order to prevent an easy detection, while collectively pushing the model's boundary toward a target instance.
Third, the samples myst be \emph{collectively inconspicuous} by bounding the collateral damage on the victim (Figure~\ref{fig:attack_illustration_testing}).
Finally, the poison samples 
must exhibit a generalized form of transferability, 
as the adversary tests the samples on a surrogate model, trained with partial knowledge along multiple dimensions, 
defined by the \textbf{FAIL} model.

By taking into account the goals, capabilities, and limitations of realistic adversaries, we also design \sysname, a targeted poisoning attack that can be applied in a broad range of settings~\footnote{Our implementation code could be found at \url{https://github.com/sdsatumd}}.
%
%
Moreover, the \sysname attack is \emph{model agnostic}: we describe concrete implementations against 4 ML systems, which use 3 different classification algorithms (convolutional neural network, linear SVM, and random forest).
The instances crafted are able to bypass three anti-poisoning defenses, including one that we adapted to account for targeted attacks. 
By subjecting \sysname to the \textbf{FAIL} analysis,
we obtain insights into the \emph{transferability of targeted poison samples}, 
and we highlight promising directions for investigating defenses against this threat. 

In summary, this paper makes three contributions:%
\begin{itemize}
\item We introduce the \textbf{FAIL} model, a general framework for modeling realistic adversaries and evaluating their impact.
The model generalizes the transferability of attacks against ML systems, across various levels of adversarial knowledge and control. 
We show that a previous black-box evasion attack is less effective
under generalized transferability.
\item We propose \sysname, a targeted poisoning attack that overcomes the limitations of prior attacks.
\sysname is effective against 4 real-world classification tasks, even when launched by a range of weaker adversaries within the \textbf{FAIL} model.
The attack also bypasses two existing anti-poisoning defenses.
\item We systematically explore realistic adversarial scenarios and the effect of partial adversary knowledge and control on the resilience of ML models against a test-time attack and a training-time attack.
Our results provide insights into the transferability of attacks across the FAIL dimensions and highlight potential directions for investigating defenses against these attacks.
\end{itemize}

This paper is organized as follows.
In Section~\ref{sec:problem} we formalize the problem and our threat model.
In Section~\ref{sec:attacker-model} we introduce the FAIL attacker model.
In Section~\ref{sec:attacks} we describe the \sysname attack and its implementation.
We present our experimental results in Section~\ref{sec:results}, review the related work in Section~\ref{sec:related-work}, and discuss the implications in Section~\ref{sec:discussion}.
%
%

\section{Problem Statement}
\label{sec:problem}

\newcommand{\reliability}[0]{\mathcal{R}\xspace}
Lack of a unifying threat model to capture the dimensions of adversarial knowledge caused existing work to diverge in terms of adversary specifications.
Prior work defined adversaries with inconsistent capabilities.
For example, in~\cite{DBLP:journals/corr/PapernotMG16} a \emph{black-box} adversary possesses knowledge of the full feature representations, whereas its counterpart in~\cite{xu2016automatically} only assumes access to the raw data (i.e. before feature extraction).

Compared to existing white-box or black-box models, in reality, things tend to be more nuanced. A commercial ML-based malware detector~\cite{kasperskylab} can rely on a publicly known architecture with proprietary data collected from end hosts, and a mixture of known features (e.g. system calls of a binary), and undisclosed features (e.g. reputation scores of the binary).
Existing adversary definitions are too rigid and cannot account for realistic adversaries against such applications. 
In this paper, we ask \textit{how can we systematically model adversaries based on realistic assumptions about their capabilities?}

Some of the recent evasion attacks~\cite{liu2016delving, DBLP:journals/corr/PapernotMG16} investigate the transferability property of their solutions.
Proven transferability increases the strength of an attack as it allows adversaries with limited knowledge or access to the victim system to craft effective instances.
Furthermore, transferability hinders defense strategies as it renders secrecy ineffective.
However, existing work generally investigates transferability under single dimensions (e.g. limiting the adversarial knowledge about the victim algorithm).
This weak notion of transferability limits the understanding of actual attack capabilities on real systems and fails to shed light on potential avenues for defenses.
This paper aims to provide a means to define and evaluate a more \emph{general transferability}, across a wide range of adversary models.
The generalized view of threat models highlights limitations of existing training-time attacks.
Existing attacks~\cite{yang2017generative,liu2017trojaning, gu2017badnets} often assume full control over the training process of victim classifiers and have similar shortcomings to white-box attacks.
Those that do not assume full control generally omit important adversarial considerations.
Targeted poisoning attacks~\cite{Nelson:2008:EML:1387709.1387716, mozaffari2015systematic, 2017arXiv171205526C} require control of the labeling process. 
However, an attacker is often unable to determine the labels assigned to the poison samples in the training set ---consider a case where a malware creator may provide a poison sample for the training set of an ML-based malware detector, but its malicious/benign label will be assigned by the engineers who train the detector.
These attacks risk being detected by existing defenses as they might craft samples that stand out from the rest of the training set.
Moreover, they also risk causing collateral damage to the classifier; for example, in Figure~\ref{fig:attack_illustration_poisoning} the attack can trigger the misclassification of additional samples from the target's true class if the boundary is not molded to include only the target.
Such collateral damage reduces the trust in the classifier's predictions, and thus the potential impact of the attack.
Therefore, we aim to observe whether an attack could address these limitations and discover \textit{how realistic is the targeted poisoning threat?}

\topic{Machine learning background}
For our purpose, a \emph{classifier} (or hypothesis) is a function $h: X\rightarrow  Y$ that maps instances to labels to perform \emph{classification}.
An \emph{instance} $\mathbf{x}\in X$ is an entity (e.g., a binary program
) that must receive a \emph{label} $y \in  Y=\{y_0, y_1,...,y_m\}$ (e.g., reflecting whether the binary is malicious
).
%
We represent an instance as a vector $\mathbf{x}=(x_1,\dots,x_n)$, where the features reflect attributes of the artifact (e.g. APIs invoked by the binary).
A function $D(\mathbf{x},\mathbf{x'})$ represents the distance in the feature space between two instances $\mathbf{x}, \mathbf{x'} \in  X$.
The function $h$ can be viewed as a separator between the malicious and benign classes in the feature space $ X$; the plane of separation between classes is called \textit{decision boundary}.
The \emph{training set} $S \subset  X$ includes instances that have known labels $Y_S \subset  Y$.
The labels for instances in $S$ are assigned using an \emph{oracle} --- for a malware classifier, an oracle could be an antivirus service such as VirusTotal, whereas for an image classifier it might be a human annotator.
%
The \emph{testing set} $T \subset  X$ includes instances for which the labels are unknown to the learning algorithm.
%

\topic{Threat model}
%
We focus on \emph{targeted poisoning attacks} against machine learning classifiers.
In this setting, we refer to the victim classifier as Alice, the owner of the target instance as Bob, and the attacker as Mallory.
Bob and Mallory could also represent the same entity.
Bob possesses an instance $\mathbf{t} \in T$ with label $y_t$, called the \textit{target},  which will get classified by Alice.
For example, Bob develops a benign application, and he ensures it is not flagged by an oracle antivirus such as VirusTotal.
Bob's expectation is that Alice would not flag the instance either.
Indeed, the target would be correctly classified by Alice after learning a hypothesis using a pristine training set $S^*$ (i.e. $h^*=A(S^*),h^*(\mathbf{t})=y_t$).
%
Mallory has partial knowledge of Alice's classifier and read-only access to the target's feature representation, but they do not control either $\mathbf{t}$ or the natural label $y_t$, which is assigned by the oracle.
Mallory pursues two goals.
The \emph{first goal} is to introduce a targeted misclassification on the target by deriving a training set $S$ from $S^*$: $h=A(S),h(\mathbf{t})=y_d$, where $y_d$ is Mallory's desired label for $\mathbf{t}$.
On binary classification, this translates to causing a false positive (FP) or false negative (FN).
An example of FP would be a benign email message that would be classified as spam, while an FN might be a malicious sample that is not detected.
In a multiclass setting, Mallory causes the target to be labeled as a class of choice.
Mallory's \emph{second goal} is to minimize the 
effect of the attack on Alice's overall classification performance.
To quantify this collateral damage, we introduce the Performance Drop Ratio (PDR), a metric that reflects the performance hit suffered by a classifier after poisoning.
This is defined as the ratio between the performance of the poisoned classifier and that of the pristine classifier: $PDR=\frac{performance(h)}{performance(h^*)}$.
The metric encodes the fact that for a low-error classifier, Mallory could afford a smaller performance drop before raising suspicions.

\section{Modeling Realistic Adversaries}
\label{sec:attacker-model}
\topic{Knowledge and Capabilities}
Realistic adversaries conducting training time or testing time attacks are constrained by an imperfect \emph{knowledge} about the model under attack and by limited \emph{capabilities} in crafting adversarial samples.
For an attack to be successful, samples crafted under these conditions must transfer to the original model.
We formalize the adversary's strength in the \textbf{FAIL} attacker model,
which describes the adversary's knowledge and capabilities along 4 dimensions:

\begin{itemize}
\item \textbf{F}eature knowledge $\mathcal{R} = \{x_{i}: x_{i} \in \mathbf{x}$, $x_{i} $ is \emph{readable}$\}$: the subset of features known to the adversary.
\item \textbf{A}lgorithm knowledge $A'$: the learning algorithm that the adversary uses to craft poison samples. 
\item \textbf{I}nstance knowledge $S'$: the labeled training instances available to the adversary.
\item \textbf{L}everage $\mathcal{W} = \{x_{i}: x_{i} \in \mathbf{x}$, $x_{i} $ is \emph{writable}$\}$: the subset of features that the adversary can modify.
\end{itemize}
The \textbf{F} and \textbf{A} dimensions constrain the attacker's understanding of the hypothesis space.
Without knowing the victim classifier $A$, the attacker would have to select an alternative learning algorithm $A'$ and hope that the evasion or poison samples crafted for models created by $A'$ transfer to models from $A$.
Similarly, if some features are unknown (i.e., partial feature knowledge), the model used for crafting instances is an approximation of the original classifier.
For classifiers that learn a representation of the input features (such as neural networks), limiting the \textbf{F} dimension results in a different, approximate internal representation that will affect the success rate of the attack.
These limitations result in an inaccurate \emph{assessment} of the impact that the crafted instances will have and affect the success rate of the attack.
%
The \textbf{I} dimension affects the accuracy of the adversary's view over the instance space.
%
As $S'$ might be a subset or an approximation of $S^*$, the poisoning and evasion samples might exploit gaps in the instance space that are not present in the victim's model.
This, in turn, could lead to an impact overestimation on the attacker side.
Finally, the \textbf{L} dimension affects the adversary's \emph{ability} to craft attack instances.
The set of modifiable features restricts the regions of the feature space where the crafted instances could lie.
For poisoning attacks, this places an upper bound on the ability of samples to shift the decision boundary while for evasion it could affect their effectiveness.
The read-only features can, in some cases, cancel out the effect of the modified ones.
An adversary with partial leverage needs extra effort, e.g. to craft more instances (for poisoning) or to attack more of the modifiable features (for both poisoning and evasion).
%

%
Prior work has investigated transferability without modeling a full range of realistic adversaries across the FAIL dimensions.
\cite{DBLP:journals/corr/PapernotMG16} focuses on the \textbf{A} dimension, and proposes a transferable evasion attack across different neural network architectures.
Transferability of poisoning samples in~\cite{munoz2017towards} is partially evaluated on the \textbf{I} and \textbf{A} dimensions.
%
%
The evasion attack in \cite{laskov2014practical} considers \textbf{F}, \textbf{A} and \textbf{I} under a coarse granularity, but omits the \textbf{L} dimension.
ML-based systems employed in the security industry~\cite{Hearn12:GoogleSpamFiltering, Chau11:Polonium, tamersoy2014guilt, CAMP, MS10:SmartScreen} often combine undisclosed and known features to render attacks more difficult.
In this context, the systematic evaluation of transferability along the \textbf{F} and \textbf{L} dimensions is still an open question.

\topic{Constraints} The attacker's strategy is also influenced by a set of \emph{constraints} that drive the attack design and implementation.
While these are attack-dependent, we broadly classify them into three categories: \textit{success}, \textit{defense}, and \textit{budget} constraints.
\textit{Success} constraints encode the attacker's goals and considerations that directly affect the effectiveness of the attack, such as the assessment of the target instance classification.
\textit{Defense} constraints refer to the attack characteristics aimed to circumvent existing defenses (e.g. the post-attack performance drop on the victim).
\textit{Budget} considerations address the limitations in an attacker's resources, such as the maximum number of poisoning instances or, for evasion attacks, the maximum number of queries to the victim model.

\topic{Implementing the FAIL dimensions}
Performing empirical evaluations within the \textbf{FAIL} model requires further design choices that depend on the application domain and the attack surface of the system.
To simulate weaker adversaries systematically, we formulate a questionnaire to guide the design of experiments focusing on each dimension of our model.

For the \textbf{F} dimension, we ask: \textit{What features could be kept as a secret?} \textit{Could the attacker access the exact feature values?}
Feature subsets may not be publicly available (e.g. derived using a proprietary malware analysis tool, such as dynamic analysis in a contained environment), or they might be directly defined from instances not available to the attacker (e.g. low-frequency word features).
Similarly, the exact feature values could be unknown (
e.g. because of defensive feature squeezing~\cite{xu2017feature}).
Feature secrecy does not, however, imply the attacker's inability to modify them through an indirect process~\cite{laskov2014practical} or extract surrogate ones.

The questions related to the \textbf{A} dimension are: \textit{Is the algorithm class known?} \textit{Is the training algorithm secret?} \textit{Are the classifier parameters secret?}
These questions define the spectrum for adversarial knowledge with respect to the learning algorithm: black-box access, if the information is public, gray-box, where the attacker has partial information about the algorithm class or the ensemble architecture, or white-box, for complete adversarial knowledge.

The \textbf{I} dimension controls the overlap between the instances available to the attacker and these used by the victim.
Thus, here we ask: \textit{Is the entire training set known?} \textit{Is the training set partially known?} \textit{Are the instances known to the attacker sufficient to train a robust classifier?}
An application might use instances from the public domain (e.g. a vulnerability exploit predictor) and the attacker could leverage them to the full extent in order to derive their attack strategy.
However, some applications, such as a malware detector, might rely on private or scarce instances that limit the attacker's knowledge of the instance space.
The scarcity of these instances drives the robustness of the attacker classifier which in turn defines the perceived attack effectiveness.
In some cases, the attacker might not have access to any of the original training instances, being forced to train a surrogate classifier on independently collected samples~\cite{xu2016automatically, liu2017trojaning}.

The \textbf{L} dimension encodes the practical capabilities of the attacker when crafting attack samples.
These are tightly linked to the attack constraints.
However, rather than being preconditions, they act as degrees of freedom on the attack.
Here we ask: \textit{Which features are modifiable by the attacker?} and \textit{What side effects do the modifications have?}
For some applications, the attacker may not be able to modify certain types of features, either because they do not control the generating process (e.g. an exploit predictor that gathers features from multiple vulnerability databases) or when the modifications would compromise the instance integrity (e.g. a watermark on images that prevents the attacker from modifying certain features).
In cases of dependence among features, targeting a specific set of features could have an indirect effect on others (e.g. an attacker injecting tweets to modify word feature distributions also changes features based on tweet counts).

\subsection{Unifying Threat Model Assumptions}

Discordant threat model definitions result in implicit assumptions about adversarial limitations, some of which might not be realistic.
The FAIL model allows us to systematically reason about such assumptions. 
To demonstrate its utility, we evaluate a body of existing studies by means of answering two questions for each work.

To categorize existing attacks, we first inspect a threat model and ask: \textit{AQ1--Are bounds for attacker limitations specified along the dimension?}.
The possible answers are: \textit{yes, omitted} and \textit{irrelevant}.
For instance, the threat model in Carlini et al.'s evasion attack~\cite{carlini2017towards} 
%
specifies that the adversary requires complete knowledge of the model and its parameters, thus the answer is \textit{yes} for the \textbf{A} dimension. 
In contrast, the analysis on the \textbf{I} dimension is \textit{irrelevant} because the attack does not require access to the victim training set.
However, the study does not discuss feature knowledge, therefore we mark the \textbf{F} dimension as \textit{omitted}.

Our second question is: \textit{AQ2--Is the proposed technique evaluated along the dimension?}.
%
This question becomes \textit{irrelevant} if the threat model specifications are \textit{omitted} or \textit{irrelevant}.
For example, Carlini et al. evaluated transferability of their attack when the attacker does not know the target model parameters.
This corresponds to the attacker algorithm knowledge, therefore the answer is \textit{yes} for the \textbf{A} dimension.
%

%
%
Applying the FAIL model reveals implicit assumptions in existing attacks.
%
An implicit assumption exists if the attack limitations are not specified along a dimension. 
%
%
Furthermore, even with explicit assumptions, some studies do not evaluate all relevant dimensions.
We present these findings about previous attacks within the FAIL model in Table~\ref{table:fail_refs_attacks}.

When looking at existing defenses through the FAIL model, we aim to observe how they achieve security: either by hiding information or limiting the attacker capabilities. 
%
For defenses that involve creating knowledge asymmetry between attackers and the defenders, i.e. secrecy, we ask: \textit{DQ1--Is the dimension employed as a mechanism for secrecy?}.
For example, feature squeezing~\cite{xu2017feature} employs feature reduction techniques unknown to the attacker; therefore the answer is \textit{yes} for the \textbf{F} dimension.

In order to identify hardening dimensions, which attempt to limit the attack capabilities, we ask: \textit{DQ2--Is the dimension employed as a mechanism for hardening?}.
For instance, the distillation defense~\cite{DBLP:conf/sp/PapernotM0JS16} against evasion modifies the neural network weights to make the attack more difficult; therefore the answer is \textit{yes} for the \textbf{A} dimension.

These defenses may come with inaccurate assessments for the adversarial capabilities and implicit assumptions.
For example, distillation limits adversaries along the \textbf{F} and \textbf{A} dimensions but employing a different attack strategy could bypass it~\cite{carlini2017towards}.
On poisoning attacks, the RONI~\cite{Nelson:2008:EML:1387709.1387716} defense assumes training set secrecy, but does not evaluate the threat posed by attackers with sufficient knowledge along the other dimensions. 
As our results will demonstrate, this implicit assumption allows attackers to bypass the defense while remaining within the secrecy bounds.

The results for the evaluated defenses are found in Table~\ref{table:fail_refs_defenses}.

\usenix{
The detailed evaluation process for each of these studies can be found in our technical report~\cite{suciu2018does}. 
}

%
%

%
\section{The \sysname Attack}

\label{sec:attacks}
Reasoning about implicit and explicit assumptions in prior defenses allows us to design algorithms which exploit their weaknesses.
In this section, we introduce \sysname, one such attack that achieves targeted poisoning while preserving overall classification performance.
\sysname is a general framework for crafting poison samples.

At a high level, our attack builds a set of poison instances by starting from base instances that are close to the target in the feature space but are labeled as the desired target label $y_d$, as illustrated in the example from Figure~\ref{fig:sample_crafting}.
Here, the adversary has created a malicious Android app $\mathbf{t}$, which includes suspicious features (e.g. the {\small\verb!WRITE_CONTACTS!} permission on the left side of the figure), and wishes to prevent a malware detector from flagging this app.
The adversary, therefore, selects a benign app $\mathbf{x_b}$ as a base instance.
To craft each poison instance, \sysname alters a subset of a base instance's features so that they resemble those of the target.
As shown on the right side of Figure~\ref{fig:sample_crafting}, these are not necessarily the most suspicious features, so that the crafted instance will likely be considered benign.
Finally, \sysname filters crafted instances based on their negative impact on instances from $S'$, ensuring that their individual effect on the target classification performance is negligible.
The sample crafting procedure is repeated until there are enough instances to trigger the misclassification of $\mathbf{t}$.
Algorithm~\ref{alg:attack_strategy} shows the pseudocode of the attack's two general-purpose procedures.

\begin{algorithm}[t]
\caption{The \sysname attack.}
\label{alg:attack_strategy}
\begin{algorithmic}[1]

\Procedure{StingRay($S',Y_{S'},\mathbf{t},y_t,y_d$)}{}
  \State $I = \emptyset$
  \State $h = A'(S')$
  \Repeat
    \State $\mathbf{x_b} = \Call{GetBaseInstance}{S',Y_{S'},\mathbf{t},y_t,y_d}$
    \State $\mathbf{x_c} = \Call{CraftInstance}{\mathbf{x_b},\mathbf{t}}$
    \If{$\Call{GetNegativeImpact}{S', \mathbf{x_c}} < \tau_{NI}$}
      \State $I = I\cup \{\mathbf{x_c}\}$
      \State $h = A'(S'\cup I)$
    \EndIf
  \Until{($|I| > N_{min}$ \textbf{and} $h(\mathbf{t}) = y_d$) \textbf{or} $|I| > N_{max}$}

  \State $PDR = \Call{GetPDR}{S',Y_{S'},I,y_d}$
  \If{$h(\mathbf{t}) \ne y_d$ \textbf{or} $PDR < \tau_{PDR}$}
    \State \textbf{return} $\emptyset$
  \EndIf
  \State \textbf{return} $I$
\EndProcedure
\Procedure{GetBaseInstance($S',Y_{S'},\mathbf{t},y_t,y_d$)}{}
  \For{$\mathbf{x_b},y_b$ \textbf{in} $\Call{Shuffle}{S',Y_{S'}}$}
    \If{$D(\mathbf{t},\mathbf{x_b}) < \tau_{D}$ \textbf{and} $y_b = y_d$}
      \State \textbf{return} $\mathbf{x_b}$
    \EndIf
  \EndFor
\EndProcedure
\end{algorithmic}
\end{algorithm}

We describe concrete implementations of our attack against four existing applications: an image recognition system, an Android malware detector, a Twitter-based exploit predictor, and a data breach predictor.
We re-implement the systems that are not publicly available, using the original classification algorithms and the original training sets to reproduce those systems as closely as possible.
In total, our applications utilize three classification algorithms---convolutional neural network, linear SVM, and random forest---that have distinct characteristics.
This spectrum illustrates the first challenge for our attack: identifying and encapsulating the application-specific steps in \sysname, to adopt a modular design with broad applicability.
Making poisoning attacks practical raises additional challenges.
For example, a na\"ive approach would be to inject the target with the desired label into the training set: $h(\mathbf{t})=y_d$ (\textbf{S.}\RNum{1}).
However, this is impractical because the adversary, under our threat model, does not control the labeling function. 
Therefore, \textsc{GetBaseInstance} works by selecting instances $\mathbf{x_b}$ that already have the desired label and are close to the target in the feature space (\textbf{S.}\RNum{2}).

A more sophisticated approach would mutate these samples and use poison instances to push the model boundary toward the target's class~\cite{mozaffari2015systematic}.
However, these instances might resemble the target class too much, and they might not receive the desired label from the oracle or even get flagged by an outlier detector.
In \textsc{CraftInstance}, we apply tiny perturbations to the instances (\textbf{D.}\RNum{3}) and by checking the negative impact $NI$ of crafted poisoning instances on the classifier (\textbf{D.}\RNum{4}) we ensure they remain \textit{individually inconspicuous}.

Mutating these instances with respect to the target~\cite{Nelson:2008:EML:1387709.1387716} (as illustrated in Figure~\ref{fig:attack_illustration_poisoning}) may still reduce the overall performance of the classifier (e.g. by causing the misclassification of additional samples similar to the target).
We overcome this via \textsc{GetPDR} by checking the performance drop of the attack samples (\textbf{S.}\RNum{5}), therefore ensuring that they remain \textit{collectively inconspicuous}.

Even so, the \sysname attack adds robustness to the poison instances by crafting more instances than necessary, to
overcome sampling-based defenses (\textbf{D.}\RNum{6}).
Nevertheless, the attack has a sampling budget that dictates the allowable number of crafted instances (\textbf{B.}\RNum{7}).
A detailed description of \sysname is found in Appendix~\ref{sec:apx-attacks}.

\topic{Attack Constraints}
The attack presented above has a series of constraints that shape its effectiveness.
Reasoning about them allows us to adapt \sysname to the specific restrictions on each application.
These span all three categories identified in Section~\ref{sec:attacker-model}: Success(\textbf{S.}), Defense(\textbf{D.}) and Budget(\textbf{B.}):

\begin{enumerate}
\renewcommand{\labelenumi}{\textbf{S.\Roman{enumi}}}
\item $h(\mathbf{t})=y_d$: the desired class label for target
\item $D(\mathbf{t},\mathbf{x_b}) < \tau_D$: the inter-instance distance metric
\renewcommand{\labelenumi}{\textbf{D.\Roman{enumi}}}
\item $\bar{s} = \frac{1}{|I|} \sum\limits_{\mathbf{x_c} \in I} s(\mathbf{x_c}, \mathbf{t})$, where $s(\cdot, \cdot)$ is a \emph{similarity} metric: crafting target resemblance
\item $NI < \tau_{NI}$: negative impact of poisoning instances
\renewcommand{\labelenumi}{\textbf{S.\Roman{enumi}}}
\item $PDR < \tau_{PDR}$: the perceived performance drop
\renewcommand{\labelenumi}{\textbf{D.\Roman{enumi}}}
\item $|I|\ge N_{min}$: the minimum number of poison instances
\renewcommand{\labelenumi}{\textbf{D.\Roman{enumi}}}
\renewcommand{\labelenumi}{\textbf{B.\Roman{enumi}}}
\item $|I| \le N_{max}$: maximum number of poisoning instances
\end{enumerate}

The perceived success of the attacker goals (\textbf{S.}\RNum{1} and \textbf{S.}\RNum{5}) dictate whether the attack is triggered.
If the $PDR$ is large, the attack might become indiscriminate and the risk of degrading the overall classifier's performance is high.
The actual $PDR$ could only be computed in the white-box setting.
For scenarios with partial knowledge, it is approximated through the perceived $PDR$ on the available classifier.

The impact of crafted instances is influenced by the distance metric and the feature space used to measure instance similarity (\textbf{S.}\RNum{2}).
For applications that learn feature representations (e.g. neural nets), the similarity of learned features might be a better choice for minimizing the crafting effort.

The set of features that are actively modified by the attacker in the crafted instances (\textbf{D.}\RNum{3}) defines the \textit{target resemblance} for the attacker, which imposes a trade-off between their inconspicuousness and the effectiveness of the sample.
If this quantity is small, the crafted instances are less likely to be perceived as outliers, but a larger number of them is required to trigger the attack.
A higher resemblance could also cause the oracle to assign crafted instances a different label than the one desired by the attacker.

The loss difference of a classifier trained with and without a crafted instance (\textbf{D.}\RNum{4}) approximates the negative impact of that instance on the classifier.
It may be easy for an attacker to craft instances with a high negative impact, but these instances may also be easy to detect using existing defenses.

In practice, the cost of injecting instances in the training set can be high (e.g. controlling a network of bots in order to send fake tweets) so the attacker aims to minimize the number of poison instances (\textbf{D.}\RNum{6}) used in the attack.
The adversary might also discard crafted instances that do not have the desired impact on the ML model.
Additionally, some poison instances might be filtered before being ingested by the victim classifier.
However, if the number of crafted instances falls below a threshold $N_{min}$, the attack will not succeed.
The maximum number of instances that can be crafted (\textbf{B.}\RNum{7}) influences the outcome of the attack. If the attacker is unable to find sufficient poison samples after crafting $N_{max}$ instances,
they might conclude that the large fraction of poison instances in the training set would trigger suspicions or that they depleted the crafting budget.
%

\topic{Delivering Poisoning Instances}
The mechanism through which poisoning instances are delivered to the victim classifier is dictated by the application characteristics and the adversarial knowledge.
In the most general scenario, the attacker injects the crafted instances alongside existing ones, expecting that the victim classifier will be trained on them.
For applications where models are updated over time or trained in mini-batches (such as an image classifier based on neural networks), the attacker only requires control over a subset of such batches and might choose to deliver poison instances through them.
In cases where the attacker is unable to create new instances (such as a vulnerability exploit predictor), they will rely on modifying the features of existing ones by poisoning the feature extraction process.
The applications we use to showcase \sysname highlight these scenarios and different attack design considerations.
%
\subsection{Bypassing Anti-Poisoning Defenses}
\label{sec:existing-defenses}
In this section, we discuss three defenses against poisoning attacks and how \sysname exploits their limitations.

\arxiv{
\vspace{0.05in}\topic{Micromodels}
}
The Micromodels defense was proposed for cleaning training data for network intrusion detectors~\cite{cretu2008casting}.
The defense trains classifiers on non-overlapping epochs of the training set (\textit{micromodels}) and evaluates them on the training set.
By using a majority voting of the micromodels, training instances are marked as either safe or suspicious. Intuition is that attacks have relatively low duration and they could only affect a few micromodels.
It also relies on the availability of accurate instance timestamps.

\arxiv{
\vspace{0.05in}\topic{RONI}
}
Reject on Negative Impact (RONI) was proposed against spam filter poisoning attacks~\cite{Barreno2010}.
It measures the incremental effect of each individual suspicious training instance and discards the ones with a relatively significant negative impact on the overall performance.
RONI sets a threshold by observing the average negative impact of each instance in the training set and flags an instance when its performance impact exceeds the threshold.
This threshold determines RONI's ultimate effectiveness and ability to identify poisoning samples. 
The defense also requires a sizable clean set for testing instances.
We adapted RONI to a more realistic scenario, assuming no clean holdout set, implementing an iterative variant, as suggested in~\cite{saini2008machine}, that incrementally decreases the allowed performance degradation threshold.
To the best of our knowledge, this version has not been implemented and evaluated before.
However, RONI remains computationally inefficient as the number of trained classifiers scales linearly with the training set.

\arxiv{
\vspace{0.05in}\topic{tRONI}
}
Target-aware RONI (tRONI) builds on the observation that RONI fails to mitigate \emph{targeted} attacks~\cite{Nelson:2008:EML:1387709.1387716} because the poison instances might not individually cause a significant performance drop.
We propose a targeted variant which leverages prior knowledge about a test-time misclassification to determine training instances that might have caused it.
While RONI estimates the negative impact of an instance on a holdout set, tRONI considers their effect on the target classification alone.
Therefore tRONI is only capable of identifying instances that distort the target classification significantly.
\usenix{A detailed description of this defense is available in the technical report~\cite{suciu2018does}.}
\arxiv{
Moreover, there exists an arms race between an attacker aware of the negative impact of poison instances (Cost \textbf{D.}\RNum{5}) and the thresholds used by tRONI, as described in Appendix~\ref{sec:apx-defenses}.
}
%
\arxiv{
\begin{algorithm}
\caption{The targeted RONI (tRONI) defense.}
\small
\label{IterativeTargetedRONI}
\begin{algorithmic}[1]
\Procedure{TargetedRONI($S,Y_{S}, \mathbf{t}, y_t,THS,n,k$)}{}
\State $R = \textit{[]}$
\ForAll{\textit{TH} \textbf{in} \textit{THS}}
\ForAll{\textit{$(x, y_{x}) \in (S, Y_{S})$}}
\If{$x \in R$}
\State \textbf{continue}
\EndIf
\State $\textit{$TS, Y_{TS}$} = $
 $\Call{GenSubsets}{S,Y_{S},R,\mathbf{t},y_t,n,k}$
\State $\textit{P} = $
 $\Call{QueryInstance}{TS,Y_{TS},x,y_{x},\mathbf{t}}$
\If{ \textit{\Call{RejectInstance}{$P,y_t,TH$}}}
\State $R = R \cup \{x\}$
\EndIf
\EndFor
\EndFor
\State \textbf{return} $R$
\EndProcedure
\end{algorithmic}
\end{algorithm}

Algorithm~\ref{IterativeTargetedRONI} describes the targeted RONI algorithm.
The \textbf{\textsc{TargetedRONI}} procedure takes as input the training set $(S,Y_{S})$, the target instance of the adversary and its true label
$(\mathbf{t}, y_t)$,
a list of significance threshold values $THS$, in decreasing order,%
the number of subsets sampled from the training set $n$,
and the size $k$ of these subsets.
We use a set $R$ to keep track of training instances that the algorithm rejects.
After $|THS|$ iterations, the algorithm returns all detected instances $R$.
For $THS$, we utilize gradual cleaning to progressively lowers the allowed
negative impact threshold for rejecting instances.
Thus, $THS$ is a decreasingly sorted list of thresholds aiming to reject
the instances starting from these having an obvious negative impact to these with subtle impact.

The \textbf{\textsc{GenSubsets}} procedure samples $n$ subsets from the $(S,Y_{S})$, each with $k$ training instances, without including the instances in $R$, which have already been discarded.
We ensure that the classifiers trained on these subsets predict the correct label $y_t$ for the target instance $\mathbf{t}$.
The procedure returns these sampled subsets and the corresponding labels, $(TS, Y_{TS})$.
%
We empirically found that $k$ should be low enough to allow individual instances to have significant impact on the classifiers trained on these subsets, allowing us to
distinguish the unreliable training instances from the reliable ones.

The \textbf{\textsc{QueryInstance}} procedure takes the subsets,
adding the currently evaluated training instance $(x,y_{x})$ to each one before
training $n$ new classifiers.
The procedure then tests these classifiers on the target instance, $\mathbf{t}$, and returns the classification results as a list of labels, $P$.

\textbf{\textsc{RejectInstance}} uses these target predictions together with
the correct target label $y_t$ and the significance threshold of current iteration $TH$.
The procedure returns $true$ if the ratio of target mispredictions to the number of classifiers $n$ exceeds the significance threshold.
In these cases, the current instance $(x, y_{x})$ is rejected.

Because the RONI variants examine training instances one by one, the order in which we iterate over the $S$ set may influence the algorithm effectiveness.
We therefore randomize the training set to avoid additional assumptions about the existing instances.
%
Moreover, adversarial samples with small \emph{individual} impact that trigger mispredictions only when \emph{collectively} used in the training process---such as the ones crafted by \sysname---may evade both RONI variants.
} 

All these defenses aim to increase adversarial costs by forcing attackers to craft instances that result in a small loss difference (Cost \textbf{D.}\RNum{4}). 
Therefore, they implicitly assume that poisoning instances stand out from the rest, and they negatively affect the victim classifier. 
However, attacks such as \sysname could exploit this assumption to evade detection by crafting a small number of inconspicuous poison samples.

\subsection{Attack Implementation}
\label{sec:attack-implementation}

We implement \sysname against four applications with distinct characteristics, each highlighting realistic constraints for the attacker.
\usenix{
We omit certain technical details for space considerations, encouraging interested readers to consult the technical report~\cite{suciu2018does}.
}
\arxiv{
  \begin{table}
    \centering
    \begin{tabular}{ll}
      \hline
      Layer Type             & Layer size            \\ \hline
      Convolution + ReLU     & $3 \times 3 \times 32$  \\
      Convolution + ReLU     & $3 \times 3 \times 64$  \\
      Max Pooling            & $2 \times 2$            \\
      Dropout                & $0.25$                  \\
      Convolution + ReLU     & $3 \times 3 \times 128$ \\
      Max Pooling            & $2 \times 2$            \\
      Convolution + ReLU     & $3 \times 3 \times 128$ \\
      Max Pooling            & $2 \times 2$            \\
      Dropout                & $0.25$                  \\
      Fully Connected + ReLU & $1024$                  \\
      Dropout                & $0.5$                   \\
      Softmax                & $10$                     \\ \hline
    \end{tabular}
  \caption{The architecture of our NN-based image classifier. }
  \label{tbl:nn-architrecutre}
  \end{table}

  \begin{table}
    \centering
    \begin{tabular}{ll}
      \hline
      Parameter     & Value \\ \hline
      Optimizer     & ADAM        \\
      Learning Rate & 0.0009      \\
      Decay         & 0.000001    \\
      Batch Size    & 32          \\
      Epochs        & 25          \\ \hline
    \end{tabular}
  \caption{The parameters of our NN-based image classifier.}
  \label{tbl:nn-params}
  \end{table}
}

  \topic{Image classification}
  We first poison a neural-network (NN) based application for image classification, often used for demonstrating evasion attacks in the prior work.
  The input instances are images and the labels correspond to objects that are depicted in the image (e.g. airplane, dog, ship).
  We evaluate \sysname on our own implementation for CIFAR-10~\cite{krizhevsky2009learning}.
  10,000 instances (1/6 of the data set) are used for validation and testing.
  In this scenario, the attacker has an image $\mathbf{t}$ with true label $y_{t}$ (e.g. a dog) and wishes to trick the model into classifying it as a specific class $y_{d}$ (e.g. a cat).

  We implement a neural network architecture that achieves a performance comparable to other studies~\cite{DBLP:conf/sp/PapernotM0JS16,carlini2017towards}, obtaining a validation accuracy of 78\%.
  \arxiv{
  Our network uses convolutional, max pooling, and dropout layers.
  The exact architecture, presented as a stack of layers, is listed in Table~\ref{tbl:nn-architrecutre}.
  For training, we used an ADAM optimizer with parameters listed in Table \ref{tbl:nn-params}.
  }
  
  Once the network is trained on the benign inputs, we proceed to poison the classifier.
  We generate and group poison instances into batches alongside benign inputs.
  We define $\gamma \in [0,1]$ to be the \emph{mixing parameter} which controls the number of poison instances in a batch.
  In our experiments we varied $\gamma$ over $\{0.125, 0.5, 1.0\}$ (i.e. 4, 16, and 32 instances of the batch are poison) and selected the value that provided the best attack success rate, keeping it fixed across successive updates.
  We then update\footnote{ The update is performed on the entire network (i.e. all layers are updated).} the previously trained network using these batches until either the attack is perceived as successful or we exceed the number of available poisoning instances, dictated by the cut-off threshold of $N_{max}$.
  It is worth noting that if the learning rate is high and the batch contains too many poison instances, the attack could become indiscriminate.
  Conversely, too few crafted instances would not succeed in changing the target prediction, so the attacker needs to control more batches.

  The main insight that motivates our method for generating adversarial samples is that there exist inputs to a network $\mathbf{x}_1, \mathbf{x}_2$ whose distance in pixel space $||\mathbf{x}_1 - \mathbf{x}_2||$ is much smaller than their distance in deep feature space $||H_i(\mathbf{x}_1) - H_i(\mathbf{x}_2)||$, where $H_i(\mathbf{x})$ is the value of the $i$\textsuperscript{th} hidden layer's activation for the input $\mathbf{x}$.
  This insight is motivated by the very existence of test-time adversarial examples, where inputs to the classifier are very similar in pixel space, but are successfully misclassified by the neural network~\cite{biggio2013evasion, szegedy2013intriguing,goodfellow2014explaining, xu2016automatically,Papernot17:BlackBoxAttacks,carlini2017towards}.
  Our attack consists of selecting \emph{base} instances that are close to the target $\mathbf{t}$ in \emph{deep feature space}, but are labeled by the oracle as the attacker's desired label $y_d$.
  \textsc{CraftInstance} creates poison images such that the \emph{distance to the target} $\mathbf{t}$ in deep feature space is minimized and the resulting adversarial image is within $\tau_{D}$ distance in pixel space to $\mathbf{t}$.
  Recent observations suggest that features in the deeper layers of neural networks are not transferable~\cite{yosinski2014transferable}.
  This suggests that the selection of the layer index $i$ in the objective function offers a trade-off between attack transferability and the magnitude of perturbations in crafted images (Cost \textbf{D.}\RNum{3}).

  \usenix{
  In our experiments we choose $H_i$ to be the third convolutional layer.
  }
  \arxiv{
  In our experiments, we chose the deep feature space to be represented by $H_3(\cdot)$, the activations of the $3$'rd hidden layer.
  This choice is a result of the \emph{transferability-perturbation} trade-off induced by the used deep feature space: selecting a lower layer in the network provides better transferability~\cite{yosinski2014transferable} of the poison instances, at the cost of increased incurred perturbation, while selecting a higher-level layer may enable instances with low perturbation in lieu of transferability.

  To generate the poison instances, we modified the JSMA method for evasion attacks proposed in~\cite{papernot2016limitations}.
  Although \sysname could technically employ any evasion attack, we picked JSMA due to its proven transferability in evasion scenarios~\cite{Papernot17:BlackBoxAttacks}.
  In JSMA, adversarial samples are generated by perturbing pairs of pixels that most increase the classification output for the attacker's target class or decrease the output for the other classes.
  This is achieved by using a \emph{Saliency Map}, which quantifies, for each input pixel, this change in the output of the network.
  The Saliency Map makes use of the \emph{Jacobian} of the function $\mathbf{F}(\cdot)$ learned by network and is expressed as:
    $$\largetriangledown \mathbf{F}(\mathbf{x}) = \frac{\partial \mathbf{F}(\mathbf{x})}{\partial \mathbf{x}}$$
  Since our objective is to minimize the deep feature distance between the input $\mathbf{x}$ and the target $\mathbf{t}$, we replace the Saliency Map with the \emph{Gradient} of the deep feature distance w.r.t. the input image.
  We can formally describe the value of the saliency map for a specific input pixel $j$ and a \emph{fixed} target $\mathbf{t}$ as:
    $$S_{\mathbf{t}}(\mathbf{x})[j] = \frac{\partial ||H_i(\mathbf{t}) - H_i(\mathbf{x})||_2}{\partial x_j}$$
  Using this formulation of the Saliency Map, we employ the optimization procedure of JSMA.
  }

  We pick 100 target instances uniformly distributed across the class labels.
  The desired label $y_d$ is the one closest to the true label $y_t$ from the \emph{attacker's classifier} point of view (i.e. it is the second best guess of the classifier).
  We set the cut-off threshold $N_{max}=64$, equivalent to two mini-batches of $32$ examples.
  The perturbation is upper-bounded at $\tau_{D} < 3.5\%$ resulting in a target resemblance $\bar{s} < 110$ pixels.
  %


\topic{Android malware detection}
The Drebin Android malware detector~\cite{arp2014drebin} uses a linear SVM classifier to predict if an application is malicious or benign.
The Drebin data set consists of 123,453 Android apps, including 5,560 malware samples.
These were labeled using 10 AV engines on VirusTotal~\cite{virustotal}, considering apps with at least two detections as malicious.
The feature space has 545,333 dimensions.
We use stratified sampling and split the data set into 60\%-40\% folds training and testing respectively, aiming to mimic the original classifier. 
Our implementation achieves 94\% F1 on the testing set.

The features are extracted from the application archives (APKs) using two techniques.
First, from the \textit{AndroidManifest} XML file, which contains meta information about the app,
the authors extract the permission requested, the application components and the registered system callbacks.
Second, after disassembling the \textit{dex} file, which contains the app bytecode, the system extracts suspicious Android framework calls, actual permission usage and hardcoded URLs.
The features are represented as bit vectors, where each index specifies whether the application contains a feature.
\begin{figure}[t]
\centering
\includegraphics[width=0.48\textwidth]{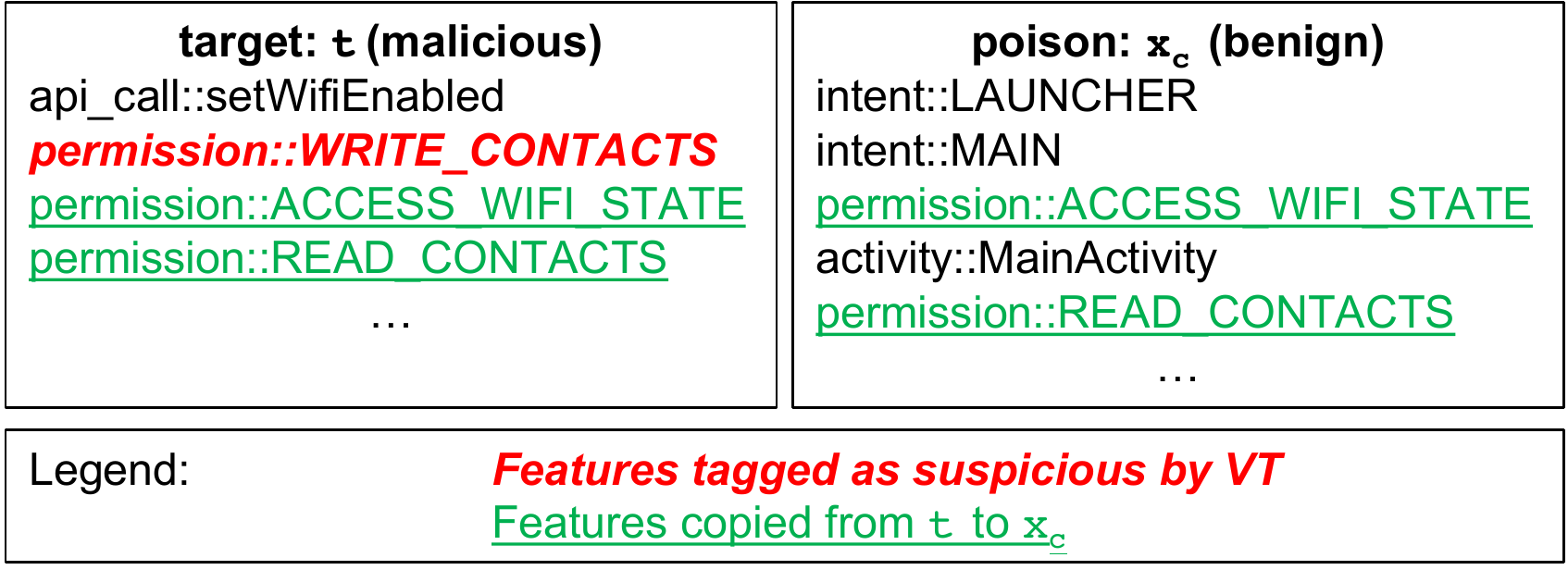}
\caption{The sample crafting process illustrated for the Drebin Android malware detector. Suspicious features are emphasized in VirusTotal reports using an opaque internal process, but the attacker is not constrained to copying them.}
\label{fig:sample_crafting}
\end{figure}
The adversary aims to misclassify an Android app $\mathbf{t}$.
Although the problems of inducing a targeted false positive (FP) and a targeted false negative (FN) are analogous from the perspective of our definitions, in practice the adversary is likely more interested in targeted FNs, so we focus on this case in our experiments.
We evaluate
this attack by selecting target instances from the testing set that would be correctly labeled as malicious by the classifier.
We then craft instances by adding active features (permissions, API calls, URL requests) from the target to existing benign instances, as illustrated in Figure~\ref{fig:sample_crafting}.
Each of the crafted apps will have a subset of the features of $\mathbf{t}$, to remain individually inconspicuous.
The poisoning instances are mixed with the pristine ones and used to train the victim classifier from scratch.

We craft 1,717 attacks to test the attack on the Drebin classifier.
We use a cutoff threshold $N_{max}=425$, which corresponds to 0.5\% of the training set.
The base instances are selected using the Manhattan distance $D = l_{1}$ and each poisoning instance has a target resemblance of $\bar{s}=10$ features and a negative impact $\tau_{NI}<50\%$.

\topic{Twitter-based exploit prediction}
In~\cite{sabottke2015vulnerability}, the authors present a system, based on a linear SVM, that predicts which vulnerabilities are going to be exploited using features extracted from Twitter and public vulnerability databases.
For each vulnerability, the predictor extracts word-based features (e.g. the number of tweets containing the word \textit{code}), Twitter statistics (e.g. number of distinct users that tweeted about it), and domain-specific features for the vulnerability (e.g. CVSS score).
%
The data set contains 4,140 instances out of which 268 are labeled as positive (a proof-of-concept exploit is publicly available).
The classifier uses 72 features from 4 categories: CVSS Score, Vulnerability Database, Twitter traffic and Twitter word features.
Due to the class imbalance, we use stratified samples of 60\%--40\% of the data set for training and testing respectively, obtaining a 40\% testing F1. 

\arxiv{
In this case, the adversary has developed a new exploit $\mathbf{t}$ for a disclosed vulnerability, and plans to weaponize it while also evading the classifier.
However, the adversary would like to evade the Twitter-based detector, to prevent potential victims from receiving an early warning about the exploitation attempts.
Because Twitter is a free and open service, the adversary may post tweets, utilize multiple accounts to emulate trending topics and may use underground services to acquire followers and trigger retweets.
However, the adversary is unable to prevent other users who observe the exploit from tweeting or to alter the vulnerability details recorded in the public vulnerability databases.
In consequence, the adversary cannot introduce new instances in the training set---unlike for the Android malware detector or the image classifier---as the vulnerabilities classified come from the vulnerability databases.
Instead, the attacker can post tweets that alter some features of vulnerabilities already included in the training set.
The targeted attack consists of
choosing a set $I$ of vulnerabilities that are similar to $\mathbf{t}$ (e.g. same product or vulnerability category), have no known exploits and gathered fewer tweets,
and
posting crafted tweets about these vulnerabilities
that include terms normally found in the tweets about the target vulnerability.
In this manner, the classifier gradually learns that these terms indicate vulnerabilities that are not exploited.
However, the attacker's capabilities are somewhat limited since the features extracted from sources other than Twitter are not under the attacker's control.
In our experiments, we select target instances for which an exploit has been reported (at a later time) in the wild or released as a proof of concept.
As part of the poisoning attack, we then choose vulnerabilities from the training set for which no known exploit has been reported.
%
%
We then attempt to match $\mathbf{t}$'s set of modifiable Twitter features ($\mathcal{W}$), simulating an attacker who controls a network of accounts with well-established statistics (e.g. the number of followers).
These accounts could originate either from black markets or compromised accounts~\cite{grier2010spam,thomas2012adapting}.
Our adversary uses these accounts to tweet about specific vulnerabilities.
In our simulation, we inject users and their tweets containing targeted keywords in the data set during the feature extraction process.
Generating tweets as a sequence of keywords is sufficient to circumvent the unigram feature model embodied by this classifier.
A more sophisticated adversary might apply natural language generation techniques~\cite{reiter2000building} to further increase the chances that posts will be deemed legitimate.
}

\usenix{
The targeted attack
selects a set $I$ of vulnerabilities that are similar to $\mathbf{t}$ (e.g. same product or vulnerability category), have no known exploits, and gathered fewer tweets.
It then proceeds to post crafted tweets about \emph{these} vulnerabilities that include terms normally found in the tweets about the \emph{target} vulnerability.
In this manner, the classifier gradually learns that these terms indicate vulnerabilities that are not exploited.
However, the attacker's leverage is limited since the features extracted from sources other than Twitter are not under the attacker's control.
}

We simulate 1,932 attacks
setting $N_{max}=20$ and selecting the CVEs to be poisoned using the Euclidean distance $D = l_{2}$ with $\tau_{NI}<50\%$.
%

\topic{Data breach prediction}
The fourth application we analyze is a data breach predictor proposed in~\cite{liu2015cloudy}.
The system attempts to predict whether an organization is going to suffer a data breach, by using a random forest classifier.
The features used in classification include indications of bad IT hygiene (e.g. misconfigured DNS servers) and malicious activity reports (e.g. blacklisting of IP addresses belonging to the organization).
These features are absolute values (i.e. organization size), as well as time series based statistics (e.g. duration of attacks).
The Data Breach Investigations Reports (DBIR)~\cite{DBIR} provides the ground truth.
The classifier uses 2,292 instances with 382 positive-labeled examples.
The 74 existing features are extracted from externally observable network misconfiguration symptoms as well as blacklisting information about hosts in an organization's network.
A similar technique is used to compute the FICO Enterprise Security Score~\cite{FICOEnterpriseSecurity}.
We use stratified sampling to build a training set containing 50\% of the corpus and use the rest for testing and choosing targets for the attacks.
The classifier achieves a 60\% F1 score on the testing set. 

In this case, the adversary plans to hack an organization $\mathbf{t}$, but wants to avoid triggering an incident prediction despite the eventual blacklisting of the organization's IPs.
In our simulation, we choose $\mathbf{t}$ from within organizations that were reported in DBIR and were not used at training time, being correctly classified at testing.
\arxiv{
To achieve their goal, the adversary chooses a set $I$ of organizations that do not appear in the DBIR and modifies their feature representation.
%
For example, the adversary could initiate malicious activities (e.g. exploitation, spam) that appear to originate from these organizations or could hack the organizations and use their servers with the sole purpose of getting them blacklisted.
%

We again assume that some of the features cannot be modified (e.g. the size of the network), while others (e.g. the blacklisting of hosts) can be manipulated indirectly through malicious activity.
This gradually forces the classifier to reduce the weight of features that the adversary cannot manipulate.

In our experiments, the attacker has limited leverage and is only able to influence time series based features indirectly, by injecting information in various blacklists.
In practice, this could be done by either compromising the blacklists or hacking the organizations themselves in order to initiate the poisoning.
}
\usenix{
The adversary chooses a set $I$ of organizations that do not appear in the DBIR and modifies their feature representation.
The attacker has limited leverage and is only able to influence time series based features indirectly, by injecting information in various blacklists.
}

We generate 2,002 attacks under two scenarios: the attacker has compromised a blacklist and is able to influence the features of many organizations, or the attacker has infiltrated a few organizations and it uses them to modify their reputation on all the blacklists.
We set $N_{max}=50$ and the instances to be poisoned are selected using the Euclidean distance $D = l_{2}$ with $\tau_{NI}<50\%$.
\subsection{Practical Considerations}
\topic{Running time of \sysname}
The main computational expenses of \sysname are: crafting the instances in \textsc{CraftInstance}, computing the distances to the target in \textsc{GetBaseInstance}, and measuring the negative impact of the crafted instances in \textsc{GetNegativeImpact}.

\textsc{CraftInstance} depends on the crafting strategy and its complexity in searching for features to perturb. 
For the image classifier, we adapt an existing evasion attack, showing that we could reduce the computational cost by finding adversarial examples on hidden layers instead of the output layer.
For all other applications we evaluated, the choice of features is determined in constant time.

The \textsc{GetBaseInstance} procedure computes inter-instance distances once per attack, and it is linear in terms of the attackers training set size for a particular label.
For larger data sets the distance computation could be approximated (e.g. using a low-rank approximation).

In \textsc{GetNegativeImpact}, we obtain a good approximation of the negative impact (NI) by training locally-accurate classifiers on small instance subsets and performing the impact test on batches of crafted instances.
%

\topic{Labeling poisoning instances}
\label{sec:labeling}
Our attacker model assumes that the adversary does not control the oracle used for labeling poisoning instances.
Although the attacker could craft poisoning instances that closely resemble the target $\mathbf{t}$ to make them more powerful, they could be flagged as outliers or the oracle could assign them a label that is detrimental for the attack.
It is therefore beneficial to reason about the oracles specific to all applications and the mechanisms used by \sysname to obtain the desired labels.

For the image classifier, the most common oracle is a consensus of human analysts.
In an attempt to map the effect of adversarial perturbations on human perception, the authors of~\cite{papernot2016limitations} found through a user study that the maximum fraction of perturbed pixels at which humans will correctly label an image is $14\%$.
We, therefore, designed our experiments to remain within these bounds.
Specifically, we measure the pixel space perturbation as the $l_{\infty}$ distance and discard poison samples with $\tau_{D} > 0.14$ prior to adding them to $I$.

The Drebin classifier uses VirusTotal as the oracle.
In our experiments, the poison instances would need to maintain the benign label.
We systematically create over 19,000 Android applications that correspond to attack instances and utilize VirusTotal, in the same way as Drebin does, to label them.
To modify selected features of the Android apps, we reverse-engineer Drebin's feature extraction process to generate apps that would have the desired feature representation.
We generate these applications for the scenario where only the subset of features extracted from the \textit{AndroidManifest} are modifiable by the attacker, similar to prior work~\cite{grosse2016adversarial}.
\arxiv{

Specifically, we start from the source code of benign applications and add nodes to the XML (e.g. new permissions) that were extracted from the target application.
We then repack these applications and submit them to VirusTotal for labeling.
We aimed to craft 25,482 applications corresponding to 549 targets.
Due to inconsistencies between the feature names and the \textit{AndroidManifest} nodes, we were unable to generate valid \textit{apk} files in 5,850 of these cases.
We then submitted the remaining 19,632 files to VirusTotal and measured their detection rate as per the Drebin paper.
For 7,785 labeled benign in the Drebin ground truth, VirusTotal returned malicious labels.
This is likely because antivirus products have started detecting them after the Drebin ground truth was labeled; incorrect labels in Drebin were also reported in prior work~\cite{zhu2016featuresmith}.
For the 11,847 benign-labeled base instances, 1,311 of their crafted instances were detected, while 10,536 passed through.
The number of crafted instances that were detected may decrease if the attacker adds fewer features to base instances (in this case, we added 10 features per instance).
%
}
In 89\% of these cases, the crafted apps bypassed detection, demonstrating the feasibility of our strategy in obtaining negatively labeled instances.
However, in our attack scenario, we assume that the attacker is not consulting the oracle, releasing all crafted instances as part of the attack.
%

%
For the exploit predictor, labeling is performed independently of the feature representations of instances used by the system.
The adversary manipulates the public discourse around existing vulnerabilities, but the label exists with respect to the availability of an exploit.
Therefore the attacker has more degrees of freedom in modifying the features of instances in $I$, knowing that their desired labels will be preserved.

In case of the data breach predictor, the attacker utilizes organizations with no known breach and aims to poison the blacklists that measure their hygiene, or hacks them directly.
In the first scenario, the attacker does not require access to an organization's networks, therefore the label will remain intact.
The second scenario would be more challenging, as the adversary would require extra capabilities to ensure they remain stealthy while conducting the attack.

%
%

\section{Evaluation}
\label{sec:results}
We start by evaluating weaker evasion and poisoning adversaries, within the \textbf{FAIL} model, on the image and malware classifiers (Section~\ref{sec:attack-effectiveness}).
Then, we evaluate the effectiveness of existing defenses against \sysname (Section~\ref{sec:existing-defense-effectiveness}) and its applicability to a larger range of classifiers.
Our evaluation seeks to answer four research questions: \textit{How could we systematically evaluate the transferability of existing evasion attacks?} \textit{What are the limitations of realistic poisoning adversaries?} \textit{When are targeted poison samples transferable?} \textit{Is \sysname effective against multiple applications and defenses?}
We quantify the effectiveness of the evasion attack using the percentage of successful attacks (SR), while for \sysname we also measure the Performance Drop Ratio (PDR).
We measure the PDR on holdout testing sets by considering either the average accuracy, on applications with balanced data sets, or the average F1 score (the harmonic mean between precision and recall), which is more appropriate for highly imbalanced data sets.
%

%
%
\subsection{FAIL Analysis}
\label{sec:attack-effectiveness}
\begin{table*}[t]
  \centering\small
\resizebox{2.1\columnwidth}{!}{
\parbox{0.30\textwidth}{
  \begin{tabular}{ | c || c | c | c ||}
  \hline
  {\small \#} & \small{$\Delta$} & \small{SR \%} & \small{$\bar{\tau_{D}}$}  \\ \hline
  \multicolumn{4}{c}{ }\\\hline
  1 & 32\%             & 67/3       & 0.070 \\  \hline
  2 & 62\%             & 86/7       & 0.054 \\  \hline

  \multicolumn{4}{c}{ }\\\hline
  3 & {\small shallow} & 99/10      & 0.035 \\  \hline
  4 & {\small narrow}  & 82/20      & 0.027 \\  \hline

  \multicolumn{4}{c}{ }\\ \hline
  5 & 35000            & 93/18      & 0.032 \\  \hline
  6 & 50000            & 80/80      & 0.026 \\  \hline

  \multicolumn{4}{c}{ }\\ \hline
  7 & 45000            & 90/18      & 0.029 \\  \hline
  8 & 50000            & 96/19      & 0.034 \\  \hline

  \multicolumn{4}{c}{ }\\ \hline
  9 & 18\%            & 80/4    	& 0.011 \\  \hline
  10 & 41\%            & 80/34    	& 0.022 \\  \hline
  11 & 62\%            & 80/80    	& 0.026 \\  \hline
  \end{tabular}
  \caption{JSMA on the image classifier}
  \label{table:attack_results_jsma}
}\parbox{0.46\textwidth}{
  \begin{tabular}{ || c | c | c | c ||}
  \hline
  \small{$\Delta$} & \small{SR \%} & \small{PDR} & \small{Instances}  \\ \hline
  \multicolumn{4}{c}{ \small \textsc{\textbf{F}ail:}\small{Unknown features} } \\ \hline
  39\%             & 87/63/67   & 0.93/0.96/0.96  & 8/4/10 \\   \hline
  66\%             & 84/71/74   & 0.94/0.95/0.95  & 8/4/9 \\   \hline

  \multicolumn{4}{c}{ \small \textsc{f\textbf{A}il:}\small{Unknown algorithm} } \\\hline
  {\small shallow} & 83/65/68 & 0.97/0.97/0.96  & 17/14/15 \\    \hline
  {\small narrow}  & 75/67/72 & 0.96/0.97/0.96  & 20/16/17 \\    \hline

  \multicolumn{4}{c}{ \small \textsc{fa\textbf{I}l:}\small{Unavailable training set} } \\ \hline
  35000            & 73/68/76   & 0.97/0.96/0.96  & 17/16/14 \\    \hline
  50000            & 78/70/74   & 0.97/0.97/0.97  & 18/16/15 \\    \hline

  \multicolumn{4}{c}{ \small \textsc{fa\textbf{I}l:}\small{Unknown training set} } \\ \hline
  45000            & 82/69/74   & 0.98/0.96/0.96  & 16/10/15 \\    \hline
  50000            & 70/62/68   & 0.95/0.96/0.96  & 17/8/17 \\   \hline

  \multicolumn{4}{c}{ \small \textsc{fai\textbf{L}:}\small{Read-only features} } \\ \hline
  25\%             & 80/70/72   & 0.97/0.97/0.97  & 19/16/15 \\    \hline
  50\%            & 80/71/76   & 0.97/0.97/0.97  & 18/16/13 \\    \hline
  75\%            & 83/78/79   & 0.97/0.97/0.96  & 16/16/12 \\    \hline
  \end{tabular}
  \caption{\sysname on the image classifier}
  \label{table:attack_results_nns}
}\parbox{0.46\textwidth}{
\begin{tabular}{ || c | c | c | c |}
  \hline
\small{$\Delta$} &  \small{SR \%} & \small{PDR} & \small{Instances}  \\ \hline
\multicolumn{4}{c}{} \\ \hline
109066 & 79/3/5 & 0.99/0.99/1.00 & 73/50/53 \\ \hline
327199 & 77/12/13 & 0.99/0.99/1.00 & 51/50/15 \\ \hline
\multicolumn{4}{c}{} \\ \hline
{\small SGD}        & 42/33/42 & 0.99/0.99/0.99 & 65/50/31 \\ \hline
{\small dSVM} & 38/35/48 & 0.99/0.99/0.99 & 78/50/61 \\ \hline
\multicolumn{4}{c}{} \\ \hline
8514  & 69/27/27 & 0.90/0.99/0.99 & 57/50/42 \\ \hline
85148 & 50/50/50 & 0.99/0.99/0.99 & 77/50/61 \\ \hline
\multicolumn{4}{c}{} \\ \hline
8514   & 53/21/24 & 0.93/0.99/1.00 & 62/50/49 \\ \hline
43865  & 36/29/39 & 1.04/0.99/0.99 & 100/50/87 \\ \hline
\multicolumn{4}{c}{} \\ \hline
851   &  73/12/13 & 0.67/0.99/1.00 & 50/50/10 \\ \hline
8514  &  49/16/17 & 0.90/0.99/1.00 & 61/50/47 \\ \hline
85148 &  32/32/32 & 0.99/0.99/0.99 & 79/50/57 \\ \hline
\end{tabular}
\caption{\sysname on the malware classifier}
\label{table:attack_results_drebin2}
}
  }
\caption*{Tables~\ref{table:attack_results_jsma}, ~\ref{table:attack_results_nns},~\ref{table:attack_results_drebin2}: FAIL analysis of the two applications. For each JSMA experiment, we report the attack SR (perceived/potential), as well as the mean perturbation $\bar{\tau_{D}}$ introduced to the evasion instances. For each \sysname experiment, we report the SR and PDR (perceived/actual/potential), as well as statistics for the crafted instances on successful attacks (mean/median/standard deviation). $\Delta$ represents the variation of the \textbf{FAIL} dimension investigated.}
\label{table:attack_results_drebin}
\end{table*}
In this section, we evaluate the image classifier and the malware detector using the \textbf{FAIL} framework.
The model allows us to utilize both a state of the art evasion attack as well as \sysname for the task.
To control for additional confounding factors when evaluating \sysname, in this analysis we purposely omit the negative impact-based pruning phase of the attack.
We chose to implement the FAIL analysis on the two applications as they do not present built-in leverage limitations and they have distinct characteristics.

\vspace{0.1in}\noindent\textbf{Evasion attack on the image classifier.}
The first attack subjected to the FAIL analysis is JSMA~\cite{papernot2016limitations}, a well-known targeted evasion attack.
Transferability of this attack has previously been studied only for an adversary with limited knowledge along the \textbf{A} and \textbf{I} dimensions~\cite{Papernot17:BlackBoxAttacks}.
We attempt to reuse an application configuration similar in prior work, implementing our own 3-layer convolutional neural network architecture for the MNIST handwritten digit data set~\cite{lecun1998mnist}.
The validation accuracy of our model is 98.95\%.
In Table~\ref{table:attack_results_jsma}, we present the average results of our 11 experiments, each involving 100 attacks.

For each experiment, the table reports the $\Delta$ variation of the \textbf{FAIL} dimension investigated, two SR statistics: \emph{perceived} (as observed by the attacker on their classifier) and \emph{potential} (the effect on the victim if all attempts are triggered by the attacker) as well as the mean perturbation $\bar{\tau_{D}}$ introduced to the evasion instances.

Experiment \#6 corresponds to the white-box adversary, where we observe that the white-box attacker could reach 80\% SR.

Experiments \#1--2 model the scenario in which the attacker has limited \textbf{Feature knowledge}.
Realistically, these scenarios can simulate an evasion or poisoning attack against a self-driving system, conducted without knowing the vehicle's camera angles---wide or narrow.
We simulate this by cropping a frame of 3 and 6 pixels from the images, decreasing the available features by 32\% and 62\%, respectively.
The attacker uses the cropped images for training and testing the classifier, as well as for crafting instances.
On the victim classifier, the cropped part of the images is added back without altering the perturbations.

With limited knowledge along this dimension (\#1-2) the perceived success remains high, but the actual SR is very low.
This suggests that the evasion attacks are very sensitive in such scenarios, highlighting a potential direction for future defenses.

We then model an attacker with limited \textbf{Algorithm knowledge}, possessing a similar architecture, but with smaller network capacity.
For the shallow network (\#3) the attacker network has one less hidden layer;
the narrow architecture (\#4) has half of the original number of neurons in the fully connected hidden layers. 
Here we observe that the shallow architecture (\#3) renders almost all attacks as successful on the attacker.
However, the potential SR on the victim is higher for the narrow setup (\#4).
This contradicts claims in prior work~\cite{Papernot17:BlackBoxAttacks}, which state that the used architecture is not a factor for success.

\noindent\textbf{Instance knowledge.}
In \#5 we simulate a scenario in which the attacker only knows 70\% of the victim training set,
while \#7-8 model an attacker with 80\% of the training set available and an additional subset of instances sampled from the same distribution.

These results might help us explain the contradiction with prior work.
Indeed, we observe that a robust attacker classifier, trained on a sizable data set, reduces the SR to 19\%, suggesting that the attack success sharply declines with fewer victim training instances available.
In contrast, in~\cite{Papernot17:BlackBoxAttacks} the SR remains at over 80\% because of the non-random data-augmentation technique used to build the attacker training set. 
As a result, the attacker model is a closer approximation of the victim one, impacting the analysis along the \textbf{A} dimension.

Experiments \#9--11 model the case where the attacker has limited \textbf{Leverage} and is unable to modify some of the instance features.
This could represent a region where watermarks are added to images to check their integrity.
We simulate it by considering a border in the image from which the modified pixels would be discarded, corresponding to the attacker being able modify to 18\%, 41\% and 62\% of an image respectively.
We observe a significant drop in transferability, although \#11 shows that the SR is not reduced with leverage above a certain threshold.


\vspace{0.1in}\noindent\textbf{\sysname on the image classifier.}
We now evaluate the poisoning attack described in~\ref{sec:attack-implementation} under the same scenarios defined above.
Table~\ref{table:attack_results_nns} summarizes our results.
%
In contrast to evasion, the table reports the SR, PDR, and the number of poison instances needed.
Here, besides the \emph{perceived} and \emph{potential} statistics, we also report the \emph{actual} SR and PDR (as reflected on the victim when triggering only the attacks perceived successful).
%

For limited \textbf{Feature knowledge}, we observe that the perceived SR is over 84\% but the actual success rate drops significantly on the victim.
However, the actual SR for \#2 is similar to the white-box attacker (\#6), showing that features derived from the exterior regions of an image are less specific to an instance.
This suggests that although reducing feature knowledge decreases the effectiveness of \sysname, the specificity of some known features may still enable successful attacks.

\begin{figure}[t]
    \begin{subfigure}[t]{0.25\textwidth}
        \centering
        \includegraphics[height=0.5in]{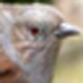}
        \includegraphics[height=0.5in]{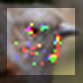}
        \includegraphics[height=0.5in]{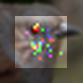}
        \caption{Limited \textbf{F}eature knowledge.}
    \end{subfigure}%
    \begin{subfigure}[t]{0.25\textwidth}
        \centering
        \includegraphics[height=0.5in]{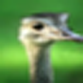}
        \includegraphics[height=0.5in]{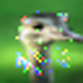}
        \includegraphics[height=0.5in]{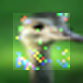}
        \caption{Limited \textbf{L}everage.}
    \end{subfigure}%
    \caption{Example of original and crafted images. Images in the left panel are crafted with 39\% and 66\% of features unknown. In the right panel, the images are crafted with 100\% and 50\% leverage.}
\label{fig:im-classif-leverage}
\end{figure}
Along the \textbf{A} dimension, we observe that both architectures allow the attacker to accurately approximate the deep space distance between instances.
While the perceived SR is overestimated, the actual SR of these attacks is comparable to the white-box attack, showing that architecture secrecy does not significantly increase the resilience against these attacks.
The open-source neural network architectures readily available for many of classification tasks would aid the adversary.
Along the \textbf{I} dimension, in \#5, the PDR is increased because the smaller available training set size prevents them from training a robust classifier.
In the white-box attack \#6 we observe that the perceived, actual and potential SRs are different.
We determined that this discrepancy is caused by documented nondeterminism in the implementation framework.
This affects the order in which instances are processed, causing variance on the model parameters, which in turns reflects on the effectiveness of poisoning instances.
Nevertheless, we observe that the potential SR is higher in \#5, even though the amount of available information is larger in \#6.
This highlights the benefit of a fine-grained analysis along all dimensions, since the attack success rate may not be monotonic in terms of knowledge levels.

Surprisingly, we observe that the actual SR for \#8, where the attacker has more training instances at their disposal, is lower than for \#7.
This is likely caused by the fact that, with a larger discrepancy between the training sets of the victim and the attacker classifier, the attacker is more likely to select base instances that would not be present in the victim training set.
After poisoning the victim, the effect of crafted instances would not be bootstrapped by the base instances, and the attacker fails.
The results suggest that the attack is sensitive to the presence of specific pristine instances in the training set, and variance in the model parameters could mitigate the threat.
However, determining which instances should be kept secret is subject for future research.


Limited \textbf{Leverage} increases the actual SR beyond the white-box attack.
When discarding modified pixels, the overall perturbation is reduced.
Thus, it is more likely that the poison samples will become collectively inconspicuous, increasing the attack effectiveness.
Figure~\ref{fig:im-classif-leverage} illustrates some images crafted by constrained adversaries.
%

The FAIL analysis results show that the perceived PDR is generally an accurate representation of the actual value, making it easy for the adversary to assess the instance inconspicuousness and indiscriminate damage caused by the attack.
The attacks transfer surprisingly well from the attacker to the victim, and a significant number of failed attacks would potentially be successful if triggered on the victim.
We observe that limited leverage allows the attacker to localize their strategy, crafting attack instances that are even \textit{more successful} than the white-box attack.

\vspace{0.1in}\noindent\textbf{\sysname on the malware classifier.}
In order to evaluate \sysname in the FAIL setting on the malware classifier, we trigger all 1,717 attacks described in~\ref{sec:attack-implementation} along 11 dimensions. 
Table~\ref{table:attack_results_drebin2} summarizes the results.
Experiment \#6 corresponds to the white-box attacker.
Experiments \#1--2 look at the case where \textbf{Features} are unknown to the adversary.
In this case, the surrogate model used to craft poison instances includes only 20\% and 60\% of the features respectively. Surprisingly, the attack is highly ineffective.
Although the attacker perceives the attack as successful in some cases, the classifier trained on the available feature subspace is a very inaccurate approximation of the original one, resulting in an actual SR of at most 12\%.
These results echo these from evasion, indicating that features secrecy might prove a viable lead towards effective defenses.
We also investigate adversaries with various degrees of knowledge about the classification \textbf{Algorithm}.
Experiment \#3 trains a linear model using the Stochastic Gradient Descent (SGD) algorithm, and in \#4 (dSVM), the hyperparameters of the SVM classifier are not known by the attacker.
Compared with the original Drebin SVM classifier, the default setting in \#4 uses a larger regularization parameter.
This suggests that regularization can help mitigate the impact of individual poison instances, but the adversary may nevertheless be successful by injecting more crafted instances in the training set.

\noindent\textbf{Instance knowledge.}
%
Experiments \#5--6 look at a scenario in which the known instances are subsets of $S^*$.
Unsurprisingly, the attack is more effective as more instances from $S^*$ become available.
The attacker's inability to train a robust surrogate classifier is reflected through the large perceived PDR.
For experiments \#7--8, victim training instances are not available to the attacker, their classifier being trained on samples from the same underlying distribution as $S^*$.
Under these constraints, the adversary could only approximate the effect of the attack on the targeted classifier.
Additionally, the training instances might be significantly different than the base instances available to the adversary, canceling the effect of crafted instances.
The results show, as in the case of the image classifier, that poison instances are highly dependent on other instances present in the training set to bootstrap their effect on target misclassification.
We further look at the impact of limited \textbf{Leverage} on the attack effectiveness.
Experiments \#9--11 look at various training set sizes for the case where only the features extracted from \textit{AndroidManifest.xml} are modifiable.
These features correspond to approximately 40\% of the 545,333 existing features.
Once again, we observe that the effectiveness of a constrained attacker is reduced.
This signals that a viable defense could be to extract features from uncorrelated sources, which would limit the leverage of such an attacker.

The FAIL analysis on the malware classifier reveals that the actual drop in performance of the attacks is insignificant on all dimensions, but the attack effectiveness is generally decreased for weaker adversaries.
However, feature secrecy and limited leverage appear to have the most significant effect on decreasing the success rate, hinting that they might be a viable defense.

\subsection{Effectiveness of \sysname}
\label{sec:existing-defense-effectiveness}
%
\begin{table}[t]
\resizebox{1.0\columnwidth}{!}{
\centering
\begin{tabular}{ | c || c || c | c | c |} \hline
  & \small{\sysname} & \small{RONI} & \small{tRONI} & \small{MM}  \\ \hline
  & \small{$|I|$/SR\%/PDR} & \multicolumn{3}{c|}{{Fix\%/PDR}} \\ \hline \hline
\small{Images} & 16/70/0.97 & -/- & -/- & -/- \\ \hline
\small{Malware} & 77/50/0.99 & 0/0.98 & 15/0.98 & -/- \\ \hline
\small{Exploits} & 7/6/1.00 & 0/0.97 & 40/0.67 & 0/0.33 \\ \hline
\small{Breach} & 18/34/0.98 & -/- & 20/0.96 & 55/0.91 \\ \hline


\end{tabular}
}
\caption{Effectiveness of \sysname and of existing defenses against it on all applications. Each attack cell reports the average number of poison instances $|I|$, the SR and actual PDR. Each defense cell reports the percentage of fixed attacks and the PDR after applying it.}
\label{table:defense_results}
\end{table}

In this section we explore the effectiveness of \sysname across all applications described in~\ref{sec:attack-implementation} and compare existing defense mechanisms in terms of their ability to prevent the targeted mispredictions. 
Table~\ref{table:defense_results} summarizes our findings.
Here we only consider the strongest (white-box) adversary to determine upper bounds for the resilience against attacks, without assuming any degree of secrecy.

\noindent \textbf{Image classifier.}
We observe that the attack is successful in 70\% of the cases and yields an average PDR of 0.97, requiring an average of 16 instances.
Upon further analysis, we discovered that the performance drop is due to other testing instances similar to the target being misclassified as $y_d$.
By tuning the attack parameters (e.g. the layer used for comparing features or the degree of allowed perturbation) to generate poison instances that are more specific to the target, the performance drop on the victim could be further reduced at the expense of requiring more poisoning instances.
Nevertheless, this shows that neural nets define a fine-grained boundary between class-targeted and instance-targeted poisoning attacks and that it is not straightforward to discover it, even with complete adversarial knowledge.

None of the three poisoning defenses are applicable on this task.
RONI and tRONI require training over 50,000 classifiers for each level of inspected negative impact.
This is prohibitive for neural networks which are known to be computationally intensive to train.
Since we could not determine reliable timestamps for the images in the data set, MM was not applicable either.

\noindent \textbf{Malware classifier.}
\sysname succeeds in half of the cases and yields a negligible performance drop on the victim.
The attack being cut off by the crafting budget on most failures (Cost \textbf{B.}\RNum{7}) suggests that some targets might be too "far" from the class separator and that moving this separator becomes difficult.
Nevertheless, understanding what causes this hardness remains an open question.

On defenses, we observe that RONI often fails to build correctly-predicting folds on Drebin and times out.
Hence, we investigate the defenses against only 97 successful attacks for which
RONI did not timeout.
MM rejects all training instances while RONI fails to detect any attack instances. 
tRONI detects very few poison instances, fixing only 15\% of attacks, as they do not have a large negative impact, individually, on the misclassification of the target.
None of these defenses are able to fix a large fraction of the induced mispredictions. 

\noindent \textbf{Exploit predictor.}
While poisoning a small number of instances, the attack has a very low success rate.
This is due to the fact that the non-Twitter features are not modifiable; if the data set does not contain other vulnerabilities similar to the target (e.g. similar product or type), the attack would need to poison more CVEs, reaching $N_{max}$ before succeeding.
The result, backed by our FAIL analysis of the other linear classifier in Section~\ref{sec:attack-effectiveness}, highlights the benefits of built-in leverage limitations in protecting against such attacks.

MM correctly identifies the crafted instances but also marks a large fraction of positively-labeled instances as suspicious.
Consequently, the PDR on the classifier is severely affected.
In instances where it does not timeout, RONI fails to mark any instance.
Interestingly, tRONI marks a small fraction of attack instances which helps correct 40\% of the predictions but still hurting the PDR.
The partial success of tRONI is due to two factors: the small number of instances used in the attack and the limited leverage for the attacker, which boosts the negative impact of attack instances through resampling.
We observed that due to variance, the negative impact computed by tRONI is larger than the one perceived by the attacker for discovered instances.
%
The adversary could adapt by increasing the confidence level of the statistic that reflects negative impact in the \sysname algorithm.
\noindent \textbf{Data breach predictor:}
The attacks for this application correspond to two scenarios, one with limited leverage over the number of time series features.
Indeed, the one in which the attacker has limited leverage has an SR of 5\%, while the other one has an SR of 63\%.
This corroborates our observation of the impact of adversarial leverage for the exploit prediction.
%
RONI fails due to consistent timeouts in training the random forest classifier.
tRONI fixes 20\% of the attacks while decreasing the PDR slightly.
MM is a natural fit for the features based on time series and is able to build more balanced voting folds.
The defense fixes 55\% of mispredictions, at the expense of lowering the PDR to 0.91.

Our results suggest that \sysname is practical against a variety of classification tasks---even with limited degrees of leverage.
Existing defenses, where applicable, are easily bypassed by lowering the required negative impact of crafted instances.
However, the reduced attack success rate on applications with limited leverage suggests new directions for future defenses.
%
%

\section{Related Work}
\label{sec:related-work}
Several studies proposed ways to model adversaries against machine learning systems.
~\cite{laskov2014practical} proposes \textit{FTC} ---\textit{features, training set}, and \textit{classifier}, a model to define an attacker's knowledge and capabilities in the case of a practical evasion attack.
Unlike the FTC model, the FAIL model is evaluated on both test- and training-time attacks, enables a fine-grained analysis of the dimensions and includes \textbf{L}everage.
These characteristics allow us to better understand how the \textbf{F} and \textbf{L} dimensions influence the attack success.
Furthermore,~\cite{liu2009game, bruckner2011stackelberg} introduce game theoretical Stackelberg formulations for the interaction between the adversary and the data miner in the case of data manipulations. 
Adversarial limitations are also discussed in~\cite{huang2011adversarial}. 
Several attacks against machine learning consider adversaries with varying degrees of knowledge, but they do not cover the whole spectrum~\cite{biggio2013evasion, papernot2016limitations, Papernot17:BlackBoxAttacks}.
Recent studies investigate transferability, in attack scenarios with limited knowledge about the target model~\cite{DBLP:journals/corr/PapernotMG16, liu2016delving, carlini2017towards}.
The FAIL model unifies these dimensions and can be used to model these capabilities systematically across multiple attacks under realistic assumptions about adversaries.
Unlike game theoretical approaches, FAIL does not assume perfect knowledge on either the attacker or the defender. 
By defining a wider spectrum of adversarial knowledge, FAIL generalizes the notion of transferability.

Prior work introduced indiscriminate and targeted poisoning attacks. 
For indiscriminate poisoning, a spammer can force a Bayesian filter to misclassify legitimate emails by including a large number of dictionary words in spam emails, causing the classifier to learn that all tokens are indicative of spam~\cite{Barreno2010}
An attacker can degrade the performance of a Twitter-based exploit predictor by posting fraudulent tweets that mimic most of the features of informative posts~\cite{sabottke2015vulnerability}. 
One could also damage the overall performance of an SVM classifier by injecting a small volume of crafted attack points~\cite{biggio2012poisoning}. 
For targeted poisoning, a spammer can trigger the filter against a specific legitimate email by crafting spam emails resembling the target~\cite{Nelson:2008:EML:1387709.1387716}.
This was also studied in the healthcare field, where an adversary can subvert the predictions for a whole target class of patients by injecting fake patient data that resembles the target class~\cite{mozaffari2015systematic}.
\sysname is a model-agnostic targeted poisoning attack and works on a broad range of applications. Unlike existing targeted poisoning attacks, \sysname aims to bound indiscriminate damage to preserve the overall performance. 

On neural networks,~\cite{koh2017understanding} proposes a targeted poisoning attack that modifies training instances which have a strong influence on the target loss.
In ~\cite{yang2017generative}, the poisoning attack is a white-box indiscriminate attack adapted from existing evasion work.
Furthermore,~\cite{liu2017trojaning} and~\cite{gu2017badnets} introduce backdoor and trojan attacks where adversaries cause the classifiers to misbehave when a trigger is present in the input.
The targeted poisoning attack proposed in~\cite{2017arXiv171205526C} requires the attacker to assign labels to crafted instances.
Unlike these attacks, \sysname does not require white-box or query access the original model. 
Our attack does not require control over the labeling function or modifications to the target instance.

%
%

\section{Discussion}
\label{sec:discussion}
The vulnerability of ML systems to evasion and poisoning attacks leads to an arms race, where defenses that seem promising are quickly thwarted by new attacks~\cite{goodfellow2014explaining,Papernot17:BlackBoxAttacks,DBLP:conf/sp/PapernotM0JS16, carlini2017towards}.
Previous defenses make implicit assumptions about how the adversary's capabilities should be constrained to improve the system's resilience to attacks. 
%
The \textbf{FAIL} adversary model provides a framework for exposing and systematizing these assumptions. 
For example, the feature squeezing defense~\cite{xu2017feature} constrains the adversary along the \textbf{A} and \textbf{F} dimensions by modifying the input features and adding an adversarial example detector.
Similarly, RONI constrains the adversary along the \textbf{I} dimension by sanitizing the training data. 
The ML-based systems employed in the security industry~\cite{Hearn12:GoogleSpamFiltering, Chau11:Polonium, CAMP, MS10:SmartScreen}, often rely on undisclosed features to render attacks more difficult, thus constraining the \textbf{F} dimension. 
In Table~\ref{table:fail_refs_defenses} we highlight implicit and explicit assumptions of previous defenses against poisoning and evasion attacks.

Through our systematic exploration of the \textbf{FAIL} dimensions, we provide the first experimental comparison of the importance of these dimensions for the adversary's goals, in the context of targeted poisoning and evasion attacks. 
For a linear classifier, our results suggest that feature secrecy is the most promising direction for achieving attack resilience.
Additionally, reducing leverage can increase the cost for the attacker.
For a neural network based image recognition system, our results suggest that \sysname's samples are transferable across all dimensions.
Interestingly, limiting the leverage causes the attacker to craft instances that are more potent in triggering the attack.
We also observed that secrecy of training instances provides limited resilience. 

Furthermore, we demonstrated that the \textbf{FAIL} adversary model is applicable to targeted evasion attacks as well.
By systemically capturing an adversary's knowledge and capabilities, the FAIL model also defines a more general notion of attack transferability.
In addition to investigating transferability under certain dimensions, such as the \textbf{A} dimension in~\cite{ carlini2017towards} or \textbf{A} and \textbf{I} dimensions in~\cite{Papernot17:BlackBoxAttacks}, generalized transferability covers a broader range of adversaries. 
At odds with the original findings in~\cite{Papernot17:BlackBoxAttacks}, our results suggest a lack of generalized-transferability for a state of the art evasion attack; while highlighting feature secrecy as the most prominent factor in reducing the attack success rate.
Future research may utilize this framework as a vehicle for reasoning about the most promising directions for defending against other attacks. 

Our results also provide new insights for the broader debate about the generalization capabilities of neural networks. 
While neural networks have dramatically reduced test-time errors for many applications, which suggests they are capable of generalization (e.g. by learning meaningful features from the training data), recent work~\cite{zhang2016understanding} has shown that neural networks can also memorize randomly-labeled training data (which lack meaningful features).
We provide a first step toward understanding the extent to which an adversary can exploit this behavior through targeted poisoning attacks.
Our results are consistent with the hypothesis that an attack, such as \sysname, can force selective memorization for a target instance while preserving the generalization capabilities of the model.
We leave testing this hypothesis rigorously for future work. 

%
%

\section{Conclusions}
\label{sec:conclusions}
We introduce the \textbf{FAIL} model, a general framework for evaluating realistic attacks against machine learning systems.
We also propose \sysname, a targeted poisoning attack designed to bypass existing defenses. 
We show that our attack is practical for 4 classification tasks, which use 3 different classifiers. 
By exploring the \textbf{FAIL} dimensions, 
we observe new transferability properties in existing targeted evasion attacks and highlight characteristics that could provide resiliency against targeted poisoning.
This exploration generalizes the prior work on attack transferability and provides new results on the transferability of poison samples. 
\paragraph{Acknowledgments}
We thank Ciprian Baetu, Jonathan Katz, Daniel Marcu, Tom Goldstein, Michael Maynord, Ali Shafahi, W. Ronny Huang, our shepherd, Patrick McDaniel and the anonymous reviewers for their feedback.
We also thank the Drebin authors for giving us access to their data set and VirusTotal for access to their service.
This research was partially supported by the Department of Defense and the Maryland Procurement Office (contract H98230-14-C-0127).




%

{\footnotesize \bibliographystyle{acm}
\bibliography{ms}}

%
%
\centerline{\Huge Appendix}
\appendix

\section{The \sysname Attack}\vspace{-0.03in}
\label{sec:apx-attacks}

Algorithm~\ref{alg:attack_strategy}
shows the pseudocode of
\sysname's two general-purpose procedures.
\textsc{StingRay} builds a set $I$ with at least $N_{min}$ and at most $N_{max}$ attack instances.
In the sample crafting loop, this procedure invokes \textsc{GetBaseInstance} to select appropriate base instances for the target.
Each iteration of the loop crafts one poison instance by invoking \textsc{CraftInstance}, which modifies the set of allowable features (according to \textbf{FAIL}'s \textbf{L} dimension) of the base instance.
This procedure is specific to each application.
The other application-specific elements are the distance function $D$ and the method for injecting the poison in the training set: the crafted instances may either replace or complement the base instances, depending on the application domain.
Next, we describe the steps that overcome the main challenges of targeted poisoning.

\topic{Application-specific instance modification}
\textsc{CraftInstance} crafts a poisoning instance by modifying the set of allowable features of the base instance.
The procedure selects a random sample among these features, under the constraint of the target resemblance budget.
It then alters these features to resemble those of the target.
Each crafted sample introduces only a small perturbation that may not be sufficient to induce the target misclassification; however, because different samples modify different features, they collectively teach the classifier that the features of $\mathbf{t}$ correspond to label $y_d$.
We discuss the implementation details of this procedure for the four applications in Section~\ref{sec:attack-implementation}.

\topic{Crafting individually inconspicuous samples}
To ensure that the attack instances do not stand out from the rest of the training set,
\textsc{GetBaseInstance} randomly selects a base instance from $S'$, labeled with the desired target class $y_d$, that lies within $\tau_D$ distance from the target.
%
By choosing base instances that are as close to the target as possible, the adversary reduces the risk that the crafted samples will become outliers in the training set.
The adversary can further reduce this risk by trading target resemblance (modifying fewer features in the crafted samples) for the need to craft more poison samples (increasing $N_{min}$).
%
The adversary then checks the negative impact of the crafted instance on the training set sample $S'$.
The crafted instance $\mathbf{x_c}$ is discarded if it changes the prediction on $\mathbf{t}$ above the attacker set threshold $\tau_{NI}$ or added to the attack set otherwise.
To validate that these techniques result in individually inconspicuous samples, we consider whether our crafted samples would be detected by three anti-poisoning defenses, discussed in detail in Section~\ref{sec:existing-defenses}.

\topic{Crafting collectively inconspicuous samples}
After the crafting stage, \textsc{GetPDR} checks the perceived $PDR$ on the available classifier.
The attack is considered successful if both adversarial goals are achieved: changing the prediction of the available classifier and not decreasing the $PDR$ below a desired threshold $\tau_{PDR}$.

\topic{Guessing the labels of the crafted samples}
By modifying only a few features in crafted sample, \textsc{CraftInstance} aims to preserve the label $y_d$ of the base instance.
While the adversary is unable to dictate how the poison samples will be labeled, they might guess this label by consulting an oracle.
We discuss the effectiveness of this technique in Section~\ref{sec:labeling}.

\arxiv{
\section{Limitations of the RONI Defense.}
\label{sec:apx-defenses}

\begin{figure}[t]
\includegraphics[width=9cm]{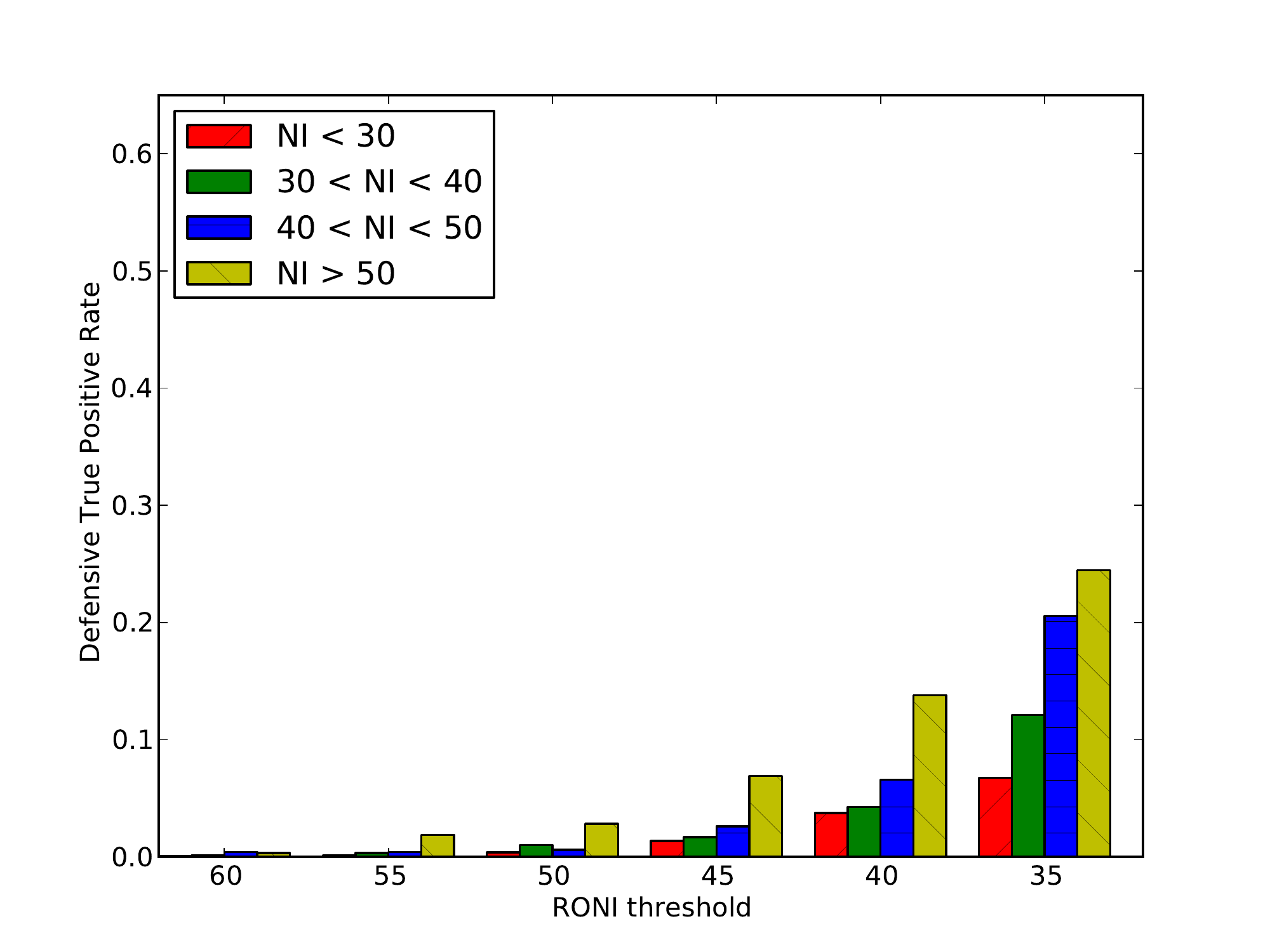}
\includegraphics[width=9cm]{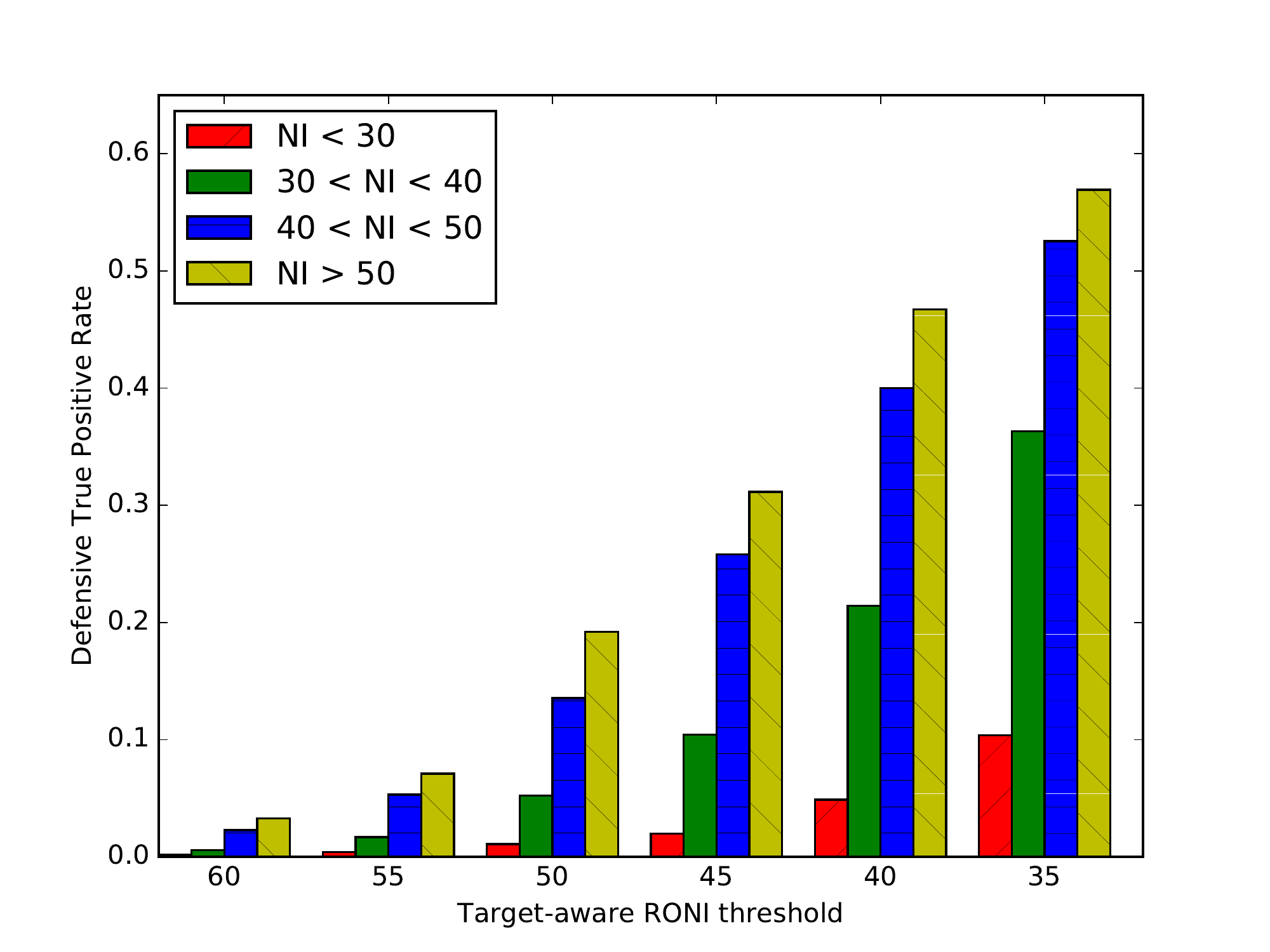}
\caption{The percentage of crafted instances detected by RONI (top) and target-aware RONI (bottom) for decreasing thresholds of negative impact. Different instance categories correspond to different attack thresholds for the negative impact $NI$.}
\label{fig:bars_tRONI}
\end{figure}

We discovered empirically that, for the best detection performance,
RONI must start by allowing a high negative impact (a higher threshold value) to reject training instances that cause the most severe performance degradation.
In subsequent iterations, this threshold must be lowered because the training set becomes cleaner with each iteration and poisoning instances have a lower negative impact.

RONI is generally effective against indiscriminate attacks, which aim to increase the false positive or false negative rates of the victim classifier.
RONI relies on the assumption that adding poisoned instances to the training set leads to performance degradation, which tends to hold for indiscriminate attacks, as the attacker aims to induce as many mispredictions as possible.

We analyze our attack's ability to bypass RONI while still inducing a misprediction on the targeted instance.

For this, we select 13, 21, and 25 targets, respectively, for the three applications to which we could apply the defense, and we craft poisoned instances for each of these targets.
Figure~\ref{fig:bars_tRONI} shows the the detection true positive rate of both RONI and tRONI on attack instances crafted against Drebin.
These instances have a variety of negative impacts $NI$, computed by the attacker during sample crafting (see Algorithm~\ref{alg:attack_strategy}), and we group them into four bins.
%
%
Both RONI variants fail to detect most of the attack instances with a low negative impact, and, as expected, tRONI has a higher detection rate than the baseline.
In some cases, RONI is able to detect some instances with an attacker-computed impact below the detection threshold.
This happens because the attacker has access to a sample of the training set, which it uses to approximate the negative impact.
This sample can sometimes underestimate the negative impact of attack instances when considered in the context of the entire training set.
Nevertheless, we observe that the number of poisoned instances that RONI can detect increases in a linear fashion as we lower the impact threshold for detection.
This illustrates the arms race between RONI and our targeted poisoning attack.
RONI variants (and tRONI in particular) can detect a larger portion of the poison samples as long as we lower the detection thresholds $THS$, which also increases the detection false positive rate.
However, the attacker can circumvent this defense by crafting instances with a smaller negative impact.
This shows that it is difficult to set the negative thresholds $THS$ manually to achieve an adequate protection against targeted poisoning attacks.

} 

\arxiv{
\clearpage
\onecolumn
\begin{longtable}{| c || c | c | c | c | p{0.5\textwidth}|}
    \hline
    Study & \textbf{F} & \textbf{A} & \textbf{I} & \textbf{L} & Explanation \\ \hline
  \hline
    \small{Genetic Evasion\cite{xu2016automatically}} & \cmark,\cmark & \cmark,\cmark  & \cmark,\xmark\mbox{\dagger}  & \cmark,\cmark   & 
\begin{itemize}
\itemsep0em
\item F: specifies that features are not fully known; also evaluates the impact of these features
\item A: the model is specified as known to the attacker, also evaluates transferability across models
\item I: requires adversary to have some benign samples to start with but does not evaluate how many, or what type of benign samples are required
\item L: specifies that some of the features cannot be modified; shows that these features are not modified in the attack
  \end{itemize}
    \\ \hline
    \small{Black-box Evasion\cite{Papernot17:BlackBoxAttacks}} & \xmark,$\emptyset$\mbox{*} & \cmark,\cmark & \cmark,\cmark  &  \xmark,$\emptyset$\mbox{*}  & 
\begin{itemize}
\itemsep0em
\item F: attacker is not specified as having full access and control of the input pixels, but the attack requires it
\item A: attacker does not know the algorithm since it assumes black-box access, also evaluates this across different surrogate and target models
\item I: attacker requires to query the model using synthetic samples, this is specified, and the number of queries is also evaluated
\item L: attacker can modify every pixel, but this is not specified
\end{itemize}
    \\ \hline
    \small{Model Stealing\cite{Tramer16:StealingMLModels}} & \cmark,\cmark & \cmark,\cmark  & \cmark,\cmark & \xmark,$\emptyset$\mbox{*}  & 
    \begin{itemize}
\itemsep0em
\item F: attacker is specified as having limited feature knowledge; attack also evaluated in a setting with partial feature knowledge
\item A: attacker is specified as knowing the model type but not knowing the model parameters; also evaluates when the model type is unknown
\item I: attacker needs to query the model, and the number of queries is evaluated too
\item L: attacker is implicitly assumed to be capable of modifying all the features in order to steal the model
\end{itemize}
    \\ \hline
    \small{FGSM Evasion\cite{goodfellow2014explaining}} & \xmark,$\emptyset$\mbox{*} & \xmark,$\emptyset$\mbox{*} & $\emptyset$,$\emptyset$ &  \xmark,$\emptyset$\mbox{*}   & 
    \begin{itemize}
\itemsep0em
\item F: attacker is not specified as knowing every pixel, but the attack requires it
\item A: attacker is not specified as knowing the victim model, and transferability is not evaluated
\item I: not applicable since knowledge of the victim model is required
\item L: attacker can modify every pixel, but this is not specified
\end{itemize}
    \\ \hline
    \small{Carlini's Evasion\cite{carlini2017towards}} & \xmark,$\emptyset$\mbox{*} & \cmark,\cmark & $\emptyset$,$\emptyset$ &  \xmark,$\emptyset$\mbox{*}  & 
    \begin{itemize}
\itemsep0em
\item F: attacker is not specified as knowing every pixel, but the attack requires it
\item A: attacker knows the model and parameters; transferability to different models is evaluated
\item I: not applicable since knowledge of the victim model is required
\item L: attacker can modify every pixel, but this is not specified
\end{itemize}
    \\ \hline
   \hline
    \small{SVM Poisoning\cite{biggio2012poisoning}} & \xmark,$\emptyset$\mbox{*} & \cmark,\xmark\mbox{\dagger}  & $\emptyset$,$\emptyset$  & \xmark,$\emptyset$\mbox{*}   & 
    \begin{itemize}
\itemsep0em
\item F: attacker implicitly assumed to know every feature and this is not evaluated
\item A: attacker is specified as knowing the model but transferability is not evaluated
\item I: not applicable since knowledge of the victim model is required
\item L: attacker can modify every feature, but this is not specified
\end{itemize}
    \\ \hline
    \small{NN Poisoning\cite{munoz2017towards}}  & \cmark,\xmark\mbox{\dagger}  & \cmark,\cmark  & \cmark,\cmark  & \cmark,\xmark\mbox{\dagger}   & 
    \begin{itemize}
\itemsep0em
\item F: attacker is specified as having full feature knowledge; but the impact of limited knowledge is not evaluated
\item A: attacker is specified as knowing the model or only a surrogate, and this is also evaluated
\item I: attacker is specified as having full or partial instance knowledge, and this is also evaluated
\item L: attacker is specified as having limited capabilities, but this is not evaluated
\end{itemize}
    \\ \hline
    \small{NN Backdoor\cite{gu2017badnets}\footnote{Gu et al.'s study investigates a scenario where the attacker performs the training on behalf of the victim. Consequently, the attacker has full access to the model architecture, parameters, training set and feature representation. However, with the emergence of frameworks such as~\cite{gilad2016cryptonets}, even in this threat model, it might be possible that the attacker does not know the training set or the features.}} & \cmark,\xmark\mbox{\dagger} & \cmark,\cmark & \cmark,\xmark\mbox{\dagger}  & \cmark,\cmark  & 
    \begin{itemize}
\itemsep0em
\item F: attacker specified as capable of changing pixels of existing instances during training, but this is not evaluated under cryptographic defense designed to hide features
\item A: attacker is specified as having full knowledge of the training algorithm and weights; attack effectiveness under different models is also evaluated
\item I: attacker requires knowledge of the training set, but this is not evaluated under cryptographic defense designed to hide instances 
\item L: specified and evaluated: full leverage during training time, but during test time attacker can only modify certain regions of an image
\end{itemize}
    \\ \hline
    \small{NN Trojan\cite{liu2017trojaning}} & \cmark,\xmark\mbox{\dagger} & \cmark, \cmark & \cmark,\cmark & \cmark,\cmark & 
    \begin{itemize}
\itemsep0em
\item F: attacker is specified as having full feature knowledge; but the impact of limited knowledge is not evaluated
\item A: attacker knows the model and parameters; transferability to different models and transfer learning is evaluated
\item I: attacker does not require knowledge of the original training instances, evaluation also considers a different training set for the attacker
\item L: attacker can only make modifications to certain region of an image at test time; the impact of these modification capabilities is also evaluated
\end{itemize}
    \\ \hline
\caption{FAIL analysis of existing attacks. For each attack, we analyze the adversary model and evaluation of the proposed technique. Each cell contains the answers to our two questions, \textit{AQ1} and \textit{AQ2}: \textit{yes} (\cmark), \textit{omitted} (\xmark) and \textit{irrelevant} ($\emptyset$). We also flag \textit{implicit assumptions} (\mbox{*}) and a \textit{missing evaluation} (\mbox{\dagger}).}
\label{table:fail_refs_attacks}
\end{longtable}%

\clearpage

\begin{longtable}{| c || c | c | c | c | p{0.5\textwidth}|}
    \hline
    Study & \textbf{F} & \textbf{A} & \textbf{I} & \textbf{L} & Explanation \\ \hline
  \hline
    \multicolumn{6}{c}{\small{Test Time Defenses} \hspace{0.3cm}} \\ \hline
    \small{Distillation\cite{DBLP:conf/sp/PapernotM0JS16}} & \xmark,\cmark  & \xmark,\cmark  & \xmark,\xmark & \xmark,\xmark & 
\begin{itemize}
\itemsep0em
\item F: hardens the learned features by changing the weights, but it does not hide these features
\item A: it does not hide the algorithm, but it hardens it by changing the weights
\item I: not used since the defense acts on the model
\item L: the defense does not influence the set of modifiable features
  \end{itemize}
    \\ \hline
    \small{Feature Squeezing\cite{xu2017feature}} & \cmark,\cmark  & \xmark,\xmark & \xmark,\xmark & \cmark,\cmark & 
\begin{itemize}
\itemsep0em
\item F: adds secret feature reduction to the inputs and also hardens the model through extra robust features
\item A: does not modify the victim model
\item I: not used since the defense acts on the model
\item L: makes some features unmodifiable by reducing the color depth, while only keeping the robust ones modifiable
  \end{itemize}
    \\ \hline
  \hline
    \multicolumn{6}{c}{\small{Training Time Defenses} \hspace{0.3cm}} \\ \hline
    \small{RONI\cite{Nelson:2008:EML:1387709.1387716}} & \xmark,\xmark  & \xmark,\xmark & \cmark,\xmark & \xmark,\xmark & 
\begin{itemize}
\itemsep0em
\item F: the defense does not influence the features
\item A: the defense does not influence the model
\item I: requires having a clean data set, or using the poisoned set to remove outliers
\item L: the defense does not influence the set of modifiable features
  \end{itemize}
    \\ \hline
    \small{Certified Defense\cite{steinhardt2017certified}} & \xmark,\xmark & \xmark,\xmark & \cmark,\cmark & \xmark,\xmark & 
\begin{itemize}
\itemsep0em
\item F: the defense does not influence the features
\item A: the defense does not influence the model
\item I: requires having a clean data set, or using the poisoned set to remove outliers and make the model more robust
\item L: the defense does not influence the set of modifiable features
  \end{itemize}
    \\ \hline
\caption{FAIL analysis of existing defenses. We analyze a defense's approach to security: \textit{DQ1} (secrecy) and \textit{DQ2} (hardening). Each cell contains the answers to the two questions: \textit{yes} (\cmark), and \textit{no} (\xmark).}
\label{table:fail_refs_defenses}
\end{longtable}

\clearpage
\twocolumn
}

\end{document}